\documentclass[prx,aps,twocolumn,nopacs,superscriptaddress,longbibliography,nofootinbib]{revtex4-1}

\usepackage{amsmath}  \usepackage{amssymb}  \usepackage{amsfonts}  \usepackage{bm}  \usepackage{bbm}   \usepackage{bbold}   \usepackage{braket}  \usepackage{color}  \usepackage{comment}  \usepackage{dcolumn}  \usepackage{enumerate}  \usepackage{epsfig}  \usepackage{gensymb}  \usepackage{graphicx}  \usepackage{indentfirst}  \usepackage{lmodern}  \usepackage{mathrsfs}  \usepackage{mathtools}  \usepackage{psfrag}  \usepackage{pst-all}  \usepackage{soul}  \usepackage{xcolor}
\usepackage{float} 
\usepackage[colorlinks,linkcolor=blue,citecolor=blue,urlcolor=blue,hyperindex,driverfallback=dvipdfm]{hyperref}  \usepackage[T1]{fontenc} 

\def\ii{{\rm i}}  \def\ee{{\rm e}}  \def\dd{{\rm d}}
\def\me{m_{\rm e}}  
  
\def\Ab{{\bf A}}        \def\Eb{{\bf E}}        \def\Gb{{\bf G}}              \def\kb{{\bf k}}            \def\qb{{\bf q}}  \def\Rb{{\bf R}}  \def\rb{{\bf r}}       
\def\xx{\hat{\bf x}}  \def\yy{\hat{\bf y}}  \def\zz{\hat{\bf z}}    \def\rr{\hat{\bf r}}        
\def\kpar{k_\parallel}  \def\kparb{{\bf k}_\parallel} 
        
  \def\qparb{{\bf q}_{\parallel}}  \def\q0parb{{\bf q}_{0\parallel}}  \def\thetae{\theta_{\rm e}}  \def\thetal{\theta_{\rm l}}
\usepackage{lipsum}
\def\neff{n_{\rm eff}}

\newcommand{\sign}{\operatorname{sign}}
\newcommand{\Hop}{\hat{\mathcal{H}}}
    
\renewcommand{\paragraph}[1]{\noindent\textbf{#1}}

\begin{document} 
\def\bibsection{\section*{\refname}} 

\title{Quantum effects in the interaction of low-energy electrons with light
}

\author{Adamantios~P.~Synanidis}
\affiliation{ICFO-Institut de Ciencies Fotoniques, The Barcelona Institute of Science and Technology, 08860 Castelldefels (Barcelona), Spain}
\author{P.~A.~D.~Gon\c{c}alves}
\affiliation{ICFO-Institut de Ciencies Fotoniques, The Barcelona Institute of Science and Technology, 08860 Castelldefels (Barcelona), Spain}
\author{Claus~Ropers}
\affiliation{Department of Ultrafast Dynamics, Max Planck Institute for Multidisciplinary Sciences, 37077 G\"ottingen, Germany}
\affiliation{4th Physical Institute - Solids and Nanostructures, University of G\"ottingen, 37077 G\"ottingen, Germany}
\author{F.~Javier~Garc\'{\i}a~de~Abajo}
\email{javier.garciadeabajo@nanophotonics.es}
\affiliation{ICFO-Institut de Ciencies Fotoniques, The Barcelona Institute of Science and Technology, 08860 Castelldefels (Barcelona), Spain}
\affiliation{ICREA-Instituci\'o Catalana de Recerca i Estudis Avan\c{c}ats, Passeig Llu\'{\i}s Companys 23, 08010 Barcelona, Spain}


\begin{abstract}{\bf \noindent
The interaction between free electrons and nanoscale optical fields has emerged as a unique platform to investigate ultrafast processes in matter and explore fundamental quantum phenomena. In particular, optically modulated electrons are employed in ultrafast electron microscopy as noninvasive probes that push the limits of spatiotemporal and spectral resolution down to the picometer--attosecond--microelectronvolt range. Electron kinetic energies well above the involved photon energies are commonly employed, rendering the electron--light coupling efficiency low and, thus, only providing limited access to the wealth of quantum nonlinear phenomena underlying the dynamical response of nanostructures. Here, we theoretically investigate electron--light interactions when photons and electrons have comparable energies, revealing strong quantum and recoil effects that include a nonvanishing coupling of surface-scattered electrons to plane waves of light, inelastic electron backscattering from localized optical fields, and strong electron--light coupling under grazing electron diffraction by an illuminated crystal surface. Our results open new vistas in electron--light--matter interactions with promising applications in ultrafast electron microscopy.
}\end{abstract}

\maketitle 

\section*{Introduction}

The synergetic relation between short light pulses and free electron beams (e-beams) underlies several recent advances in ultrafast electron microscopy toward a combined sub-nm--sub-fs--sub-meV spatiotemporal and spectral resolution, rapidly progressing toward the goal of mapping atomic-scale spatial features and their evolution over unprecedentedly small time scales~\cite{BFZ09,FZ12,FES15,BZ15,paper338,paper371,CKY23,paper415}. A prominent example is photon-induced near-field electron microscopy~\cite{BFZ09,paper151,PLZ10} (PINEM), which is based on the synchronous arrival of laser and electron femtosecond pulses at a sampled nanostructure, thus enabling optical-pump/electron-probe spectroscopy to be performed with a nanoscale spatial resolution inherited from the use of state-of-the-art electron optics setups. This approach has been applied to image optical near fields in nanophotonics~\cite{PLQ15,paper282,VMC20,KDS21}, the subcycle evolution of those fields~\cite{NKS23,GLS24,BNS23}, and the nanoscale-resolved fluctuations of the light with which the electron has interacted~\cite{paper339,DGH21,WHR24}. PINEM can be regarded as a specific application of the more general concept of stimulated inelastic electron--light scattering (SIELS). The latter has been leveraged to gain control over the free-electron wave function by customizing its interaction with light, including the generation of trains of attosecond electron pulses~\cite{FES15,PRY17,KSH18,MB18_2} and the shaping of the transverse e-beam profile~\cite{paper332,paper351,paper368,paper397,MWS22}. Laser-assisted photoemission~\cite{GSC96,SMA08} constitutes another example of SIELS that can be used to probe the ultrafast dynamics of condensed matter systems~\cite{AOM16} and produces interesting effects in the strong-field limit~\cite{YGR11,DPV20}.

Many of these advances have been accomplished in transmission electron microscopes (TEMs) operating with relatively high e-beam energies ($\gtrsim30\,$keV), orders of magnitude larger than those of the employed photons (typically in the eV range) and, consequently, rendering the probability that a single electron interacts with a single photon (e.g., in an individual confined optical mode) much smaller than unity. Such a weak electron--light interaction limits applications in metrology, imaging of atomic-scale excitations, and the study of nonlinear phenomena. For robust structures, the problem is circumvented in SIELS by using intense laser pulses~\cite{PLQ15,paper282,VMC20}, but this approach cannot be extended to sensitive specimens such as biological materials. Phase matching between the electron excitation and the light field can also boost the interaction~\cite{paper180,KLS20,WDS20,DNS20,HRF21}, although this strategy is only practical in specialized structures that host modes with evanescent optical fields in vacuum.

Surprisingly, in the linear electron--photon interaction regime, describing electrons as classical point-like charges produces the same results as a quantum-mechanical treatment in which electron recoil is ignored~\cite{paper371}. Likewise, the wave function of energetic electrons in PINEM, and more generally in SIELS, is modified by a global factor that encapsulates the interaction with light through a single complex parameter that depends linearly on the optical field~\cite{VD1971,WHC1977,paper151,PLZ10,FES15,paper371} and is therefore unsuited to access the subcycle nonlinear dynamics of a specimen in general. Consequently, it is highly desirable to achieve strong interaction between single electrons and atomic-scale excitations as a way to access their ultrafast nonlinear dynamics. In this context, the use of low-energy electrons with kinetic energies comparable to those of the quanta associated with the employed optical fields opens a plausible avenue to overcome these challenges. Indeed, the electron--light coupling strength increases when the electron energy is reduced, while a deviation from a classical-probe behavior is introduced by electron recoil~\cite{T18,T20,paper393}. Low-energy electrons thus hold the potential to reveal new physical phenomena during their interaction with localized optical fields, a possibility that demands the exploration of physically relevant configurations.

\begin{figure*}[htb] 
\centering\includegraphics[width=0.70\textwidth]{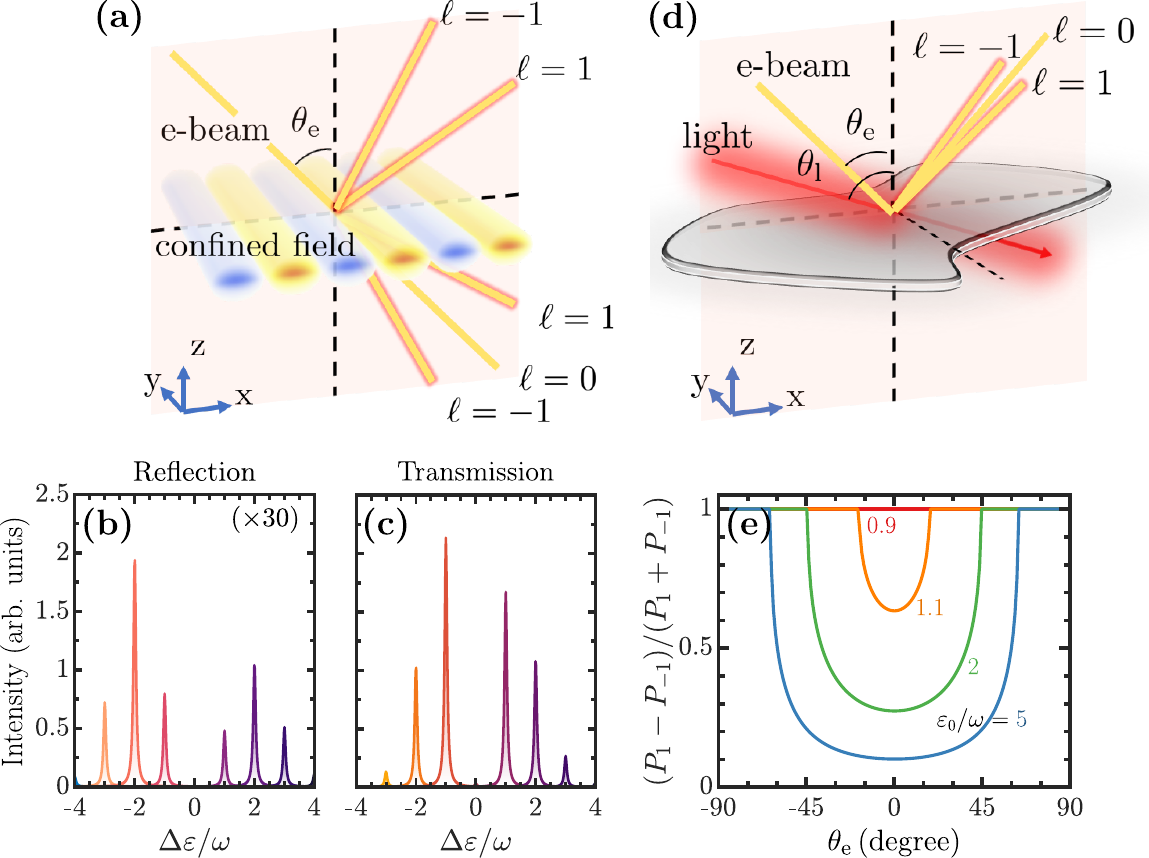}
\caption{\textbf{Electron--light--matter interaction and recoil effects with low-energy electrons.} ({\bf a})~Illustration of inelastic electron scattering by the evanescent optical field associated with propagating surface polaritons. ({\bf b}, {\bf c})~Transmission (b) and reflection (c) electron spectra in the configuration of (a) for electrons of incident energy $\hbar\varepsilon_0=10\,$eV and angle $\thetae=45\degree$ combined with polaritons of electric-field amplitude $E_0=6\times10^8\,$V/m, effective refractive index $\neff=50$, and energy $\hbar\omega=1\,$eV. For readability, spectral peaks are broadened with a Lorentzian of 0.1 FWHM in $\Delta\varepsilon/\omega$, with $\Delta\varepsilon=\varepsilon-\varepsilon_0$. ({\bf d})~Nonvanishing interaction between a plane wave of light and an electron reflected on a light-transparent surface. ({\bf e})~Despite the small coupling in (d) ($P_1<2\%$), strong recoil effects are observed for $\varepsilon_0\sim\omega$ in the asymmetry factor $(P_{1}-P_{-1})/(P_{1}+P_{-1})$, which vanishes in the classical regime ($\varepsilon_0\gg\omega$); we take $E_0=8\times10^7\,$V/m, $\thetae=45\degree$, $\thetal=90\degree$, and $\hbar\omega=1\,$eV.}
\label{Fig1}
\end{figure*}

Here, we reveal a wealth of previously unexplored quantum and recoil effects taking place during electron--light--matter interactions when the electron and photon energies are comparable. Considering realistic frameworks that admit rigorous semi-analytical treatments, we theoretically explore electron--light interactions mediated by scattering at planar interfaces of either the light, the electrons, or both (Fig.~\ref{Fig1}). We adopt optical fields in the form of either externally incident plane waves or surface polaritons. Importantly, surface scattering leads to symmetry breaking that enhances the electron--light coupling, even allowing otherwise forbidden electron--photon interactions. In particular, we show that low-energy electrons can be inelastically scattered solely due to an evanescent optical field propagating along an electron-transparent surface (Fig.~\ref{Fig1}a--c), including the emergence of a back-reflected electron signal (Fig.~\ref{Fig1}b). In addition, we demonstrate that a plane-wave electron reflected on a light-transparent surface can produce inelastically reflected electrons because of the nonvanishing electron--photon coupling originating in translational symmetry breaking of the out-of-plane electron wave function (Fig.~\ref{Fig1}d--e). The resulting reflected-electron spectrum exhibits substantial recoil effects (Fig.~\ref{Fig1}e). We further report on the possibility of reaching the strong electron--light coupling regime with moderate light intensities through Bragg scattering at planar atomic lattices, whereby the interaction is boosted under Rayleigh anomaly conditions\cite{R1907-2}. Besides its interest from a fundamental viewpoint, the present study unveils exciting opportunities for the exploration of improved microscopy and metrology in the regime of low-energy electrons exposed to optical fields of comparable photon energy. Our results are particularly relevant to the exploration of the rich phase and electronic-structure phenomenology exhibited by material surfaces, which are cornerstones in many technological applications.

\section*{Results} 

\subsection*{Theoretical framework for the interaction of low-energy electrons with light}

We first consider a planar material acting through a one-dimensional (1D) potential $V(z)$ on the electron, while a generalization to laterally corrugated atomic lattices is presented further below. To study electron scattering by the planar structure in the presence of a classical optical field (e.g., light plane waves or surface polaritons), we write the Hamiltonian $\hat{\mathcal{H}}_0(\rb)+\hat{\mathcal{H}}_1(\rb,t)$, where $\hat{\mathcal{H}}_0(\rb)=-\hbar^2\nabla^2/2\me+V(z)$ describes the electron--material system and $\hat{\mathcal{H}}_1(\rb,t)=-(\ii e \hbar/\me c)\,\Ab(\rb,t)\cdot\bm{\nabla}$ accounts for the electron--light interaction. The latter arises from the minimal coupling prescription applied to a classical vector potential $\Ab(\rb,t)$ after neglecting $A^2$ terms and adopting a gauge in which the scalar potential is zero.

We focus on monochromatic fields of frequency $\omega$ and in-plane wave vector $\kparb=(k_x,k_y)$, characterized by a vector potential $\Ab(\rb,t)=\Ab(z)\,\ee^{\ii\kparb\cdot\Rb-\ii\omega t}+{\rm c.c.}$, where the notation $\Rb=(x,y)$ is adopted. Also, in the absence of illumination, the electron wave function is taken to be a solution of the $\hat{\mathcal{H}}_0(\rb)$ Hamiltonian, namely, $\psi_0(\rb,t)=\varphi_{00}(z) \, \ee^{\ii\q0parb\cdot\Rb-\ii\varepsilon_0 t}$, having a well-defined energy $\hbar \varepsilon_0$ and in-plane wave vector $\q0parb$ ($\perp\zz$). Now, in-plane translation symmetry and energy conservation allow us to write the perturbation series
\begin{align}
\psi(\rb,t) = \sum_{n=0}^\infty \sum_{\ell=-n}^n \varphi_{n\ell}(z) \, \ee^{ \ii (\q0parb + \ell \,\kparb) \cdot \Rb-\ii (\varepsilon_0 + \ell \omega) t}, \label{psisum}
\end{align}
where $n$ runs over scattering orders, while $\ell$ denotes the net number of exchanged photons (i.e., absorbed or emitted by the electron for $\ell>0$ and $\ell<0$, respectively). In addition, $\psi(\rb,t)$ satisfies the Lippmann--Schwinger equation $\psi(\rb,t)=\psi_0(\rb,t)+\int\dd^3\rb'\int\dd t'\,\mathcal{G}_0(\rb,\rb',t-t')\,\hat{\mathcal{H}}_1(\rb',t)\,\psi(\rb',t')$, where the Green function $\mathcal{G}_0(\rb,\rb',t-t')$ is defined by $[\hat{\mathcal{H}}_0(\rb)-\ii\hbar\partial_t]\mathcal{G}_0(\rb,\rb',t-t')=-\delta(\rb-\rb')\delta(t-t')$. Combining these elements, we obtain the recurrence relation
\begin{align}
&\varphi_{n\ell}(z)=\frac{\hbar e}{\me c} \int \dd{z'}\, \mathcal{G}_0(z,z',\varepsilon^\perp_\ell) \label{eq:LPE_psi_nl}\\
&\times\Big\{
\Ab(z') \cdot \big[\q0parb + (\ell-1)\,\kparb -\ii \zz \, \partial_{z'} \big] \,\varphi_{n-1,\ell-1}(z') \nonumber\\
&\;\;\, +\Ab^*(z') \cdot \big[\q0parb + (\ell+1)\,\kparb -\ii \zz \, \partial_{z'} \big] \,\varphi_{n-1,\ell+1}(z')
\Big\} \nonumber
\end{align}
for $n>0$, where $\mathcal{G}_0(z,z',\varepsilon^\perp_\ell)$ is the frequency-domain 1D Green function satisfying $\mathcal{G}_0(\rb,\rb',t-t')=(2\pi)^{-3}\int\dd^2\qparb\int \dd\varepsilon\,\mathcal{G}_0(z,z',\varepsilon)\,\ee^{\ii\qparb\cdot(\Rb-\Rb')-\ii\varepsilon(t-t')}$, and we define $\varepsilon^\perp_\ell = \varepsilon_0 + \ell \omega - \hbar |\q0parb + \ell\, \kparb|^2/2\me$ such that $\hbar\varepsilon^\perp_\ell$ is the out-of-plane electron energy after the exchange of a net number of photons $\ell$ (see Appendix for a self-contained derivation).

We consider planar structures of negligible thickness and expand the electron wave function components in Eq.~(\ref{psisum}) using the ansatz
\begin{align}
\varphi_{n\ell}(z)=\sum_{s=\pm} \sum_j \alpha_{n\ell}^{js} \, \ee^{\zeta_{n\ell}^{js} z}\,\Theta(sz)\label{eq:phi_nlj_decomposition}
\end{align}
within the regions above ($s=+$) and below ($s=-$) the material. Here, $j$ labels contributions that can be either evanescent (${\rm Re}\{\zeta_{n\ell}^{j\pm}\}\neq0$) or propagating (imaginary $\zeta_{n\ell}^{j\pm}$). Inserting Eq.~\eqref{eq:phi_nlj_decomposition} into Eq.~\eqref{eq:LPE_psi_nl}, we find a recursive expression with a unique solution for the coefficients $\alpha_{n\ell}^{j\pm}$ and $\zeta_{n\ell}^{j\pm}$, which automatically satisfy the physical conditions ${\rm Re}\{\zeta_{n\ell}^{j+}\}\le0$ and ${\rm Re}\{\zeta_{n\ell}^{j-}\}\ge0$ (see SI). In addition, the purely propagating components have exponential coefficients $\zeta_{n\ell}^{j\pm}=\pm\ii q_{\ell z}$ (i.e., imaginary and independent of $n$ and $j$), where $q_{\ell z}=\sqrt{2\me\varepsilon^\perp_\ell/\hbar}$ is determined by energy conservation for a net number of photon exchanges $\ell$. Finally, the fractions of $\ell$-resolved electrons scattered along the upward ($+$) and downward ($-$) directions are given by
\begin{align}
P_{\ell}^\pm =\frac{q_{\ell z}}{q_{0z}}\, \Big| \sideset{}{'}\sum_{nj} \alpha_{n\ell}^{j\pm} \Big|^2, \label{eq:P_l}
\end{align}
where the primed sum indicates that it is restricted to purely propagating waves.

\begin{figure*}[ht!]
\centering\includegraphics[width=0.9\textwidth]{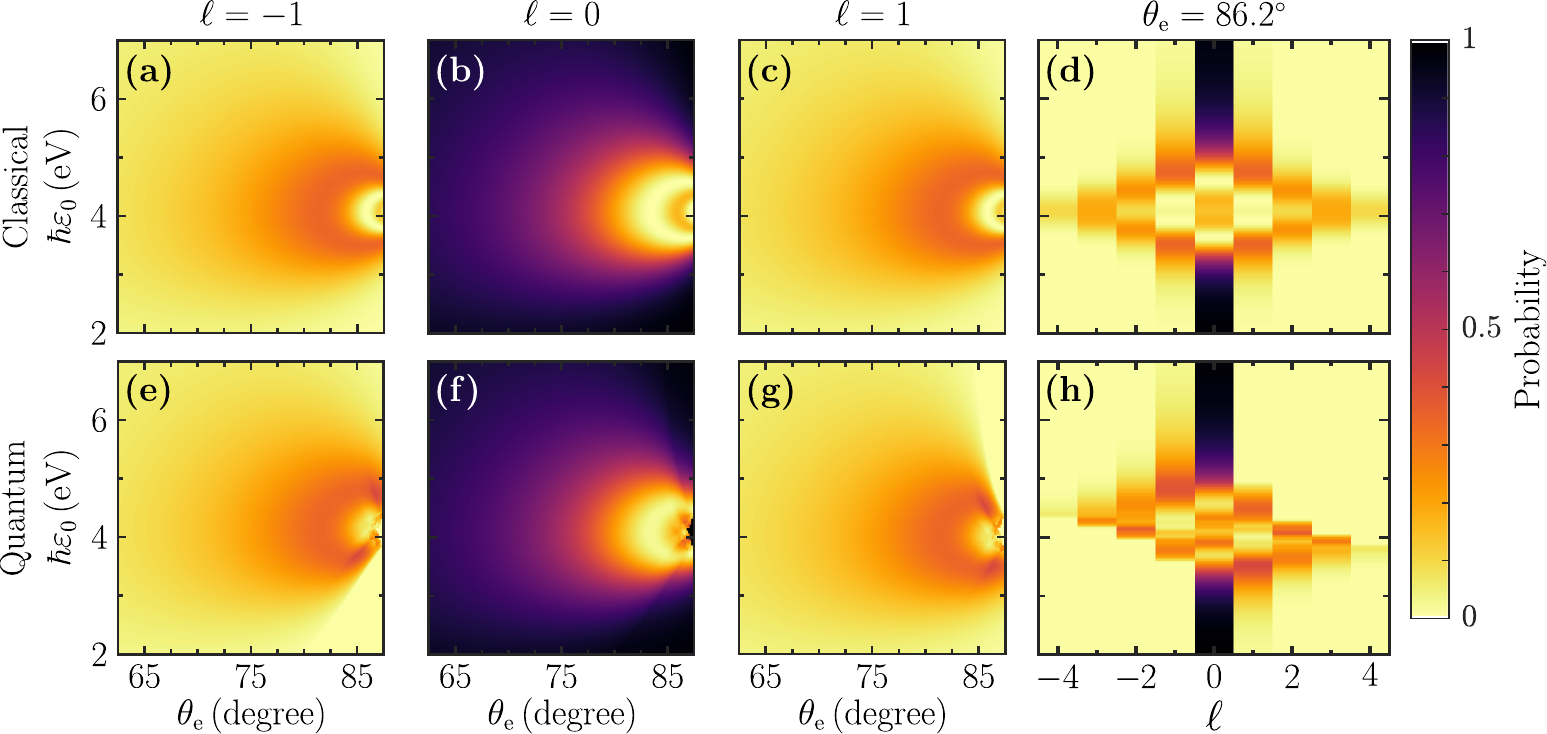}
\caption{\textbf{Inelastic scattering of low-energy electrons upon total reflection at a polariton-supporting surface.} ({\bf a--c},{\bf e--g})~Probabilities corresponding to a net exchange of $\ell={0,\pm 1}$ quanta calculated without (a--c) and with (e--g) inclusion of quantum recoil as a function of electron incidence angle $\thetae$. ({\bf d}, {\bf h})~Electron spectra for $\thetae = 86.2\degree$ obtained without (d) and with (h) recoil. We consider a polaritonic electric-field amplitude $E_0=5\times10^6\,$V/m, effective refractive index $\neff=250$, and energy $\hbar\omega=0.2\,$eV.}
\label{Fig2}
\end{figure*}

\subsection*{Recoil and quantum effects in surface-scattered electrons}

{\it Fully electron-reflecting surface.---}As a tutorial configuration, we first consider an electron with energy $\hbar\varepsilon_0$ and incident angle $\thetae$ (with respect to the surface normal) that undergoes total reflection at a planar surface supporting a surface polariton with in-plane wave vector $\kpar = |\kparb| = \neff k$, where $k=\omega/c$ is the light wave vector and $\neff>1$ is an effective index of refraction. For a thin metallic film of permittivity $\epsilon<0$ and thickness $d$ embedded in a medium of permittivity $\epsilon_s>1$, we have $\neff=\epsilon_s\lambda/[\pi(1-\epsilon)d]$ at a light wavelength $\lambda$, so thin films favor large values of $\neff$ (e.g., $\neff=40$ for $\lambda=500$~nm in currently available~\cite{paper335} 10-monolayer crystalline Ag(111) films deposited on Si). Higher values of $\neff\sim100$'s are displayed by infrared graphene plasmons~\cite{WLG15,NMS18} and phonon-polaritons in few-layer hBN~\cite{GDV18,paper361}.

For simplicity, we take the electron and the surface polariton to share the same in-plane direction of incidence with $\q0parb \parallel \kparb \parallel \xx$, such that the associated vector potential can be written as $\Ab(\rb)=(E_0/k)\,\big(\kappa^2+\kpar^2\big)^{-1/2}\,\big(\kappa\,\xx+\ii\,\kpar\,\sign\{z\}\,\zz\big)\,\ee^{\ii\kpar x-\kappa|z|}$, where $E_0$ is a global electric-field amplitude (see SI). In addition, the electron--surface interaction is assumed to be elastic, and thus, any inelastic electron signal stems from surface-polariton emission and absorption processes by the electron. In this scenario, we can set the $z$ component of the zeroth-order electron wave function as $\varphi_{00}(z) = \big[\ee^{-\ii q_{0z} z} - e^{\ii q_{0z} z} \big] \Theta(z)$, while the Green function in Eq.~\eqref{eq:LPE_psi_nl} reduces to $\mathcal{G}_0(z,z',\varepsilon^\perp_\ell)=(\ii\me/\hbar^2 q_{\ell z})\big[ \ee^{\ii q_{\ell z}(z+z')} - \ee^{\ii q_{\ell z}|z-z'|}\big] \Theta(z) \Theta(z')$ (see SI). Inserting these elements into Eq.~\eqref{eq:LPE_psi_nl} and noticing that only reflected components need to be considered, we find a set of analytical coefficients $\alpha_{n,\ell}^{j+}$ and $\zeta_{n,\ell}^{j+}$, from which the reflection probability $P_\ell\equiv P^+_\ell$ for a given $\ell$ channel is obtained via Eq.~\eqref{eq:P_l}.

It is instructive to examine the $\varepsilon_0 \gg \omega$ limit, where recoil effects should play a minor role. As a direct generalization of the result obtained for an electron moving with constant velocity along a straight-line trajectory~\cite{paper371}, we approximate the $\ell$-dependent inelastic probability as  $P_\ell = J_\ell^2(2|\beta|)$, where $\beta=(\ii e/\hbar c)\int\dd{t}\;\dot{\rb}_{\rm e}(t)\cdot \Ab[\rb_{\rm e}(t)]\, \ee^{-\ii\omega t}$ is an electron--light coupling parameter obtained by integrating over time the vector potential component parallel to the velocity $\dot{\rb}_{\rm e}(t)$ and evaluated at the electron position $\rb_{\rm e}(t)$. Taking a specularly reflected trajectory with the surface-polariton field given above, we obtain
\begin{align}
 \beta = \frac{2\,\ii}{\sqrt{2\neff^2-1}}\;\frac{eE_0c}{\hbar\omega^2}  \frac{(\neff c/v - \sin\thetae)\cos\thetae}{(c/v - \neff \sin \thetae)^2 + (\neff^2-1)\cos^2\thetae}. \label{classical}
\end{align}
For reference, we note that the prefactor $eE_0c/\hbar\omega^2$ takes a value of $\approx2$ for $E_0=10^7$~V/m and $\hbar\omega=1$~eV. This mode energy is characteristic of surface plasmons in ultrathin metal films~\cite{paper335} and exciton-polaritons in transition-metal dichalcogenides~\cite{LCZ14,ETC20}, while polaritons of lower energy (e.g., $\hbar\omega\sim0.1$~eV in graphene~\cite{WLG15,NMS18} and hBN~\cite{GDV18,paper361}) should produce larger coupling for the same field amplitude in accordance to the scaling $\beta\propto1/\omega^2$. Equation~(\ref{classical}) reveals the important role of confinement in enhancing the electron--polariton coupling: given a certain electron velocity $v$, the denominator reaches its minimum value under the condition
\begin{align}
c/\neff=v\sin\thetae \label{phasematching}
\end{align}
(i.e., when the surface-polariton phase velocity matches the in-plane projection of the electron velocity). In addition, Eq.~(\ref{classical}) illustrates the well-known linear scaling of the coupling coefficient with the applied electric field amplitude $E_0$.

Figure~\ref{Fig2} highlights the importance of recoil effects in the interaction between a low-energy electron and a strongly confined surface polariton by contrasting the classical nonrecoil PINEM theory [Fig.~\ref{Fig2}a--d, based on Eq.~(\ref{classical})] with the rigorous quantum formalism introduced above [Fig.~\ref{Fig2}e--h, Eq.~(\ref{eq:P_l})]. As a first observation, the classical treatment in Eq.~(\ref{classical}) provides the necessary conditions for strong coupling and leads to substantial inelastic probabilities (Fig.~\ref{Fig2}a-d). Focusing for concreteness on the $\ell = \pm 1$ channels (see SI for more sidebands), the parameter space for which electron--light coupling is maximized is well captured by the classical framework, and it corresponds to the phase-matching condition in Eq.~(\ref{phasematching}) (i.e., $\hbar\varepsilon_0\approx\me c^2/2\neff^2\sim4$\,eV with $\thetae=90\degree$), for which the coupling diverges as $\beta\propto 1/\cos{\theta_\mathrm{e}}$ near $\thetae=90\degree$. However, both the intensity profile in the $(\varepsilon_0,\thetae)$ phase space and the magnitude of the electron--light coupling strength are markedly different in the classical and quantum theories. Specifically, the incorporation of recoil in the latter leads to asymmetric loss--gain spectra (i.e., $P_{-\ell} \neq P_\ell$) as well as abrupt thresholds in $P_\ell$ (Fig.~\ref{Fig2}e--g). These observations can be interpreted by noting that energy--momentum conservation imposes the condition $q_{\ell z}= q_0 \sqrt{1 + \ell\omega/\varepsilon_0 - |\sin\thetae + \ell \kpar/q_0|^2}$, and thus, not only the emission ($\ell < 0$) and absorption ($\ell > 0$) probabilities are rendered different, but also the inelastic signal vanishes whenever $1 + \ell \omega/\varepsilon_0 < |\sin\thetae + \ell \kpar/q_0|^2$, since $q_{\ell z}$ becomes purely imaginary (i.e., the corresponding electron wave is evanescent). Another dramatic consequence of recoil is the redistribution of probability to neighboring $\ell$'s near the aforementioned thresholds (see Fig.~\ref{Fig2}h), in contrast to the symmetric spectrum produced by the classical description (Fig.~\ref{Fig2}d).

\begin{figure*}[ht!]
\centering\includegraphics[width=0.9\textwidth]{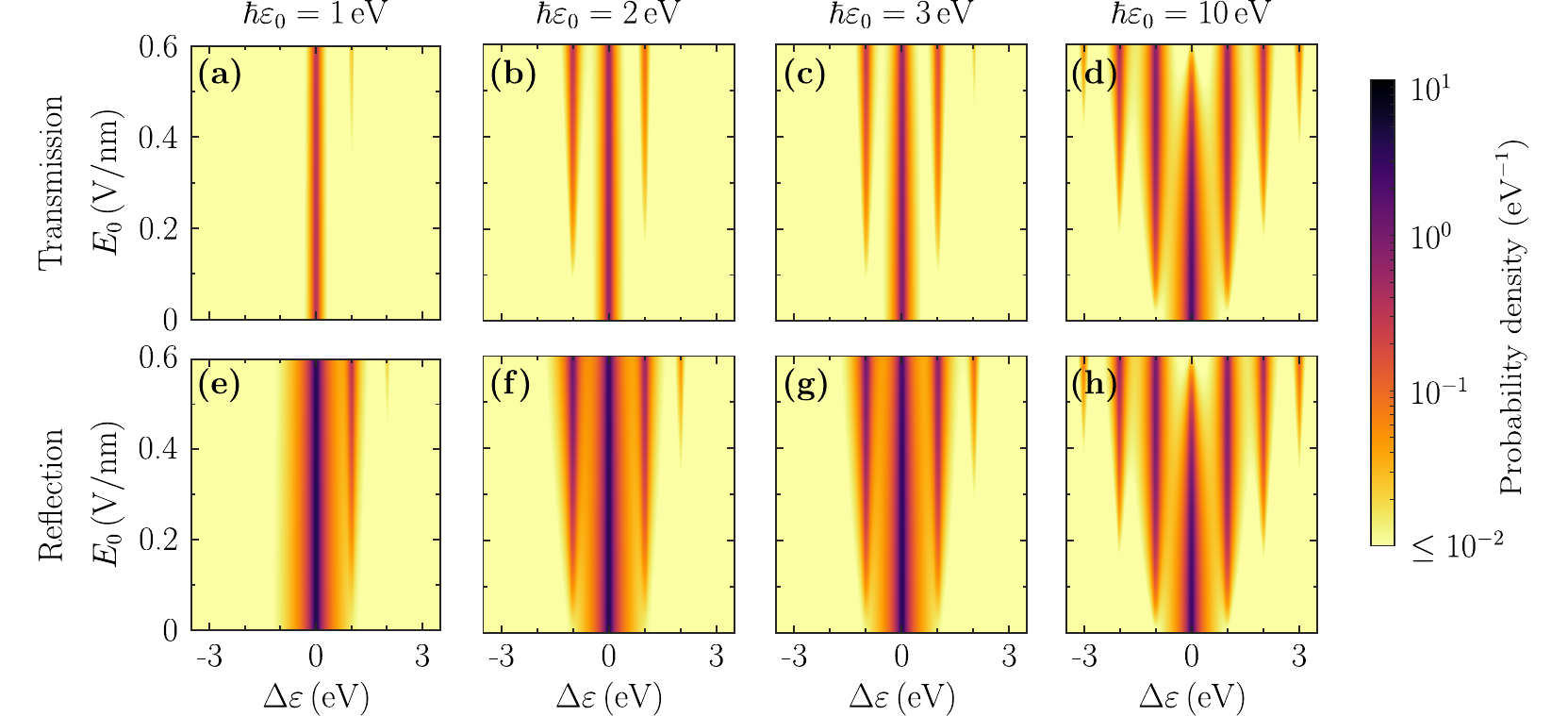}
\caption{\textbf{Inelastic scattering of low-energy electrons upon partial reflection at a polariton-supporting thin film.} We show transmitted (a--d) and reflected (e-h) electron spectra as a function of polaritonic electric-field amplitude $E_0$ for fixed effective refractive index ($\neff=50$) and energy ($\hbar\omega=1\,$eV). Selected incident electron energies $\hbar\varepsilon_0$ are considered, while the incidence angle is fixed at $\thetae=45\degree$. Electron--surface scattering is modeled through a $\delta$-function potential of amplitude $U_0=1\,$eV\,nm. Spectral features are broadened by a Lorentzian of 0.1~eV FWHM.}
\label{Fig3}
\end{figure*}

{\it Partially electron-reflecting surface.---}We expect partial transmission and reflection when the electron is scattered by an atomically thin 2D material, which we describe through a surface potential $V(z)=U_0\,\delta(z)$. The parameter $U_0$ has units of energy times length, and arguing that an atomic monolayer can be described by a barrier of finite thickness $d\lesssim1$~nm and internal potential $V_0$ in the eV range, we expect $U_0\approx V_0d$ in the eV$\times$nm range. We also assume the material to support long-lived polaritonic modes (e.g., phonon-polaritons in hBN~\cite{GDV18} or plasmons in doped graphene~\cite{NMS18}), so that they are characterized by a real effective index $\neff$. The calculation of the probabilities associated with the electron wave functions involved in the net exchange of $\ell$ polariton quanta follows the same steps as in the above scenario of full reflection, but now both reflected and transmitted electron components are produced. For an electron prepared with an incident $-q_{0z}$ wave vector in the out-of-plane direction, the wave function in the absence of illumination is given by $\varphi_{00}(z) = \big[ \ee^{-\ii q_{0z}z} + r_0 \, \ee^{\ii q_{0z}z}\big] \Theta (z) + t_0 \, \ee^{-\ii q_{0z}z}\, \Theta (-z)$, where we use the transmission and reflection coefficients $t_\ell = \big(1+\ii\me U_0/\hbar^2 q_{\ell z}\big)^{-1}$ and $r_\ell = t_\ell-1$, respectively. We use this result together with the Green function $\mathcal{G}_0(z,z',\varepsilon^\perp_\ell) = -(\ii\me/\hbar^2 q_{\ell z}) \Big[\ee^{\ii q_{\ell z}|z-z'|} + r_\ell\, \ee^{\ii q_{\ell z}(|z|+|z'|)}\Big]$ (see SI) to obtain the reflected and transmitted $\ell$-resolved probabilities $P^\pm_\ell$ from Eq.~(\ref{eq:P_l}) following the formalism developed above. The results are plotted as a function of electric-field amplitude for different electron energies in Fig.~\ref{Fig3}. Again, we find recoil effects emerging through strong asymmetries in the electron spectra, which, as anticipated, become more symmetric as the electron energy is increased toward the $\varepsilon_0\gg\omega$ regime. These results are qualitatively correct even when more involved $z$-dependent potentials are considered (e.g., finite-thickness films) because the Green function outside the material retains the same expression as above~\cite{A93}, with reflection and transmission coefficients depending on the details of the potential.
 
\subsection*{Nonvanishing interaction of surface-scattered electrons and unscattered light plane waves}

An interesting scenario is presented when a thin film is illuminated from the far field and the scattered optical components are comparatively negligible, so the electron mainly sees an external light plane wave. Electron--light coupling can still take place under this configuration because the free-space energy--momentum mismatch is broken by the fact that the electron is scattered by the material. We regard this situation as the complementary of electron shaping mediated by PINEM interaction (i.e., SIELS assisted by electron-transparent, light-reflecting plates, in which the kinematic electron--photon free-space coupling mismatch is circumvented by having light half-plane waves instead of full plane waves~\cite{paper311}). We note that 2D monolayers (e.g., graphene and hBN) are nearly transparent to light (e.g., $\sim 2.3\%$ absorption by graphene over a wide spectral range~\cite{MSW08,NBG08}) and can thus be regarded as good candidates to explore the interaction of light plane waves with surface-reflected low-energy electrons.

\begin{figure}[h!] 
\centering\includegraphics[width=0.45\textwidth]{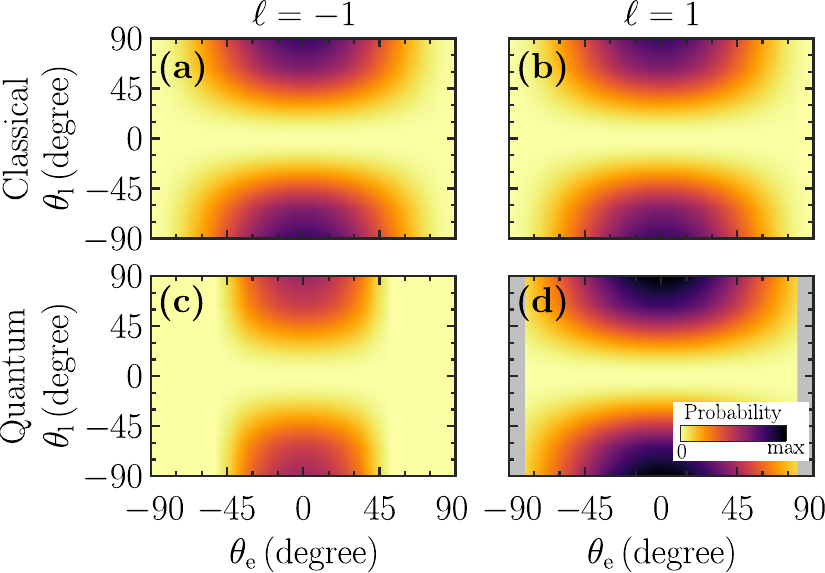}
\caption{\textbf{Inelastic interaction between a light plane wave and surface-scattered electrons.} We plot the probability associated with $\ell = \pm 1$ net photon exchanges as a function of the electron ($\thetae$) and photon ($\thetal$) incidence angles without (a,b) and with (c,d) inclusion of recoil for an electron-opaque, light-transparent surface.  We assume total-electron reflection and take $\hbar\varepsilon_0=2\,$eV, $E_0=8\times10^7\,$V/m, and $\hbar\omega=1\,$eV. In panel (d), kinematically forbidden regions are shaded in gray.}
\label{Fig4}
\end{figure}

We explore this idea in Fig.~\ref{Fig4}, where a fully reflected electron is considered to be interacting with a freely propagating p-polarized light plane wave. The analysis is analogous to that in Fig.~\ref{Fig2}, but using a plane-wave optical field instead of a surface mode. We present the probabilities $P_{\ell=\pm 1}^\pm$ calculated to first-order in the light intensity as a function of light and electron incidence angles, comparing classical (nonrecoil) and quantum (Eq.~(\ref{eq:P_l}), with recoil) descriptions. Incidentally, the coupling parameter in the former is given by
\begin{align}
\beta=\frac{2e E_0 c}{\hbar \omega^2}\frac{\cos\thetae(\sin\thetal\;c/v-\sin{\thetae})}{(c/v-\sin\thetal \sin \thetae)^2 -\cos^2\thetal\cos^2\thetae}, \nonumber
\end{align}
where $E_0$ is the light-plane-wave amplitude. We find again that recoil leads to asymmetric inelastic electron signals, as well as regions of the $(\thetal,\thetae)$ parameter space in which electron--light coupling becomes kinematically allowed or forbidden, accompanied by a transfer of probability to the {\it symmetric} ($\ell \to -\ell$) channel. In contrast to the interaction with surface polaritons, where phase-matching at grazing incidence produced the strongest interaction, now phase-matching is forbidden (i.e., the in-plane optical wave vector is always smaller than the in-plane electron wave vector), rendering the coupling smaller, so we reduce the analysis to first-order processes. The maximum coupling is observed at grazing light incidence ($\thetal=\pm90\degree$) and normal electron incidence ($\thetae=0$), which is consistent with the angular scaling of the classical coupling coefficient as $\beta\propto\sin\theta_l\cos\thetae$ for $v\ll c$. These conditions guarantee maximum overlap of the light electric field along the electron trajectory.

\begin{figure*}[ht!]
\centering\includegraphics[width=1.0\textwidth]{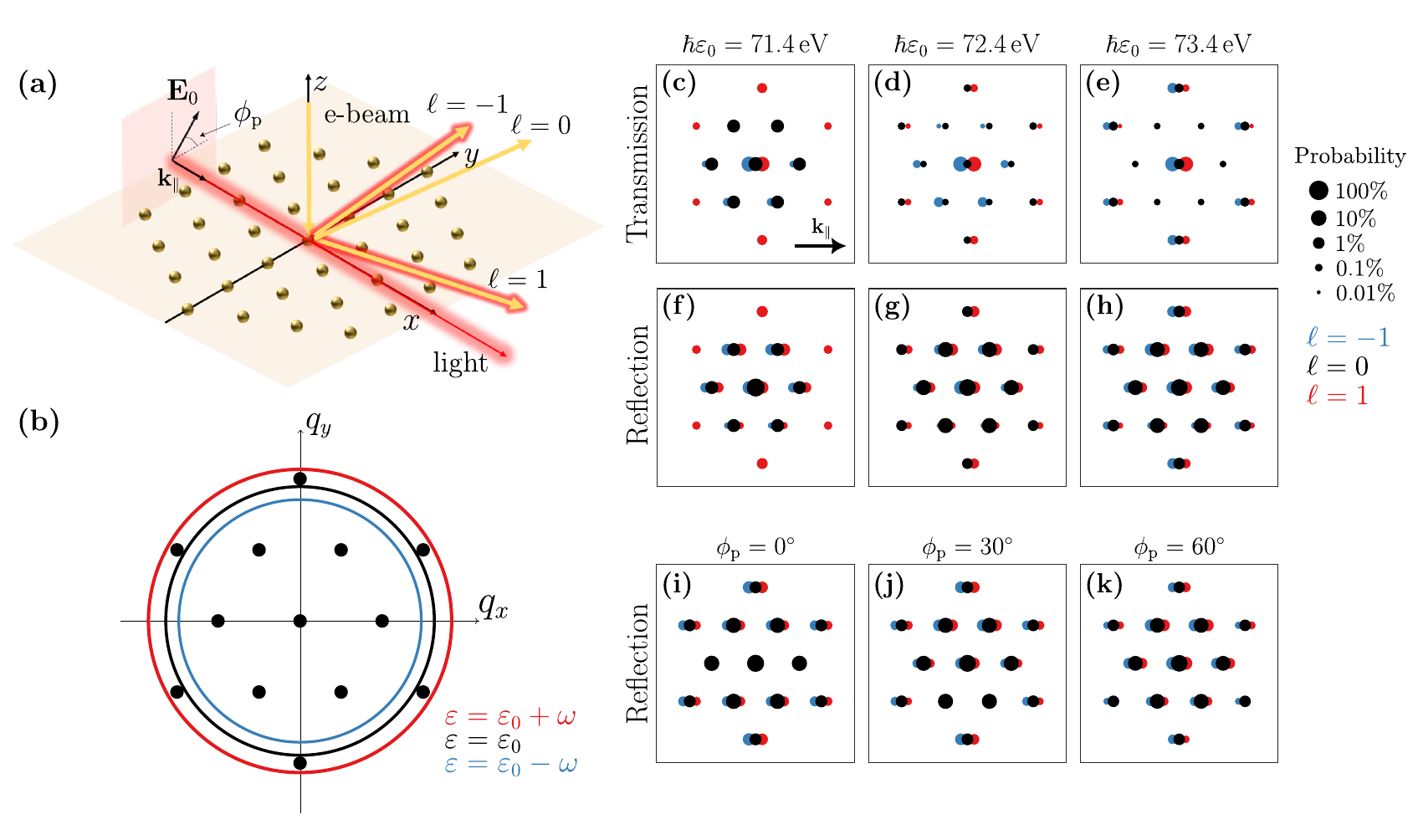}
\caption{{\bf Recoil effects in the interaction of light with lattice-diffracted electrons.} ({\bf a})~We consider a normally incident electron undergoing diffraction by an Au(111) monolayer (Au--Au distance $d\approx0.288$~nm, with atomic bonds along $y$) under grazing plane-wave light irradiation along $x$ ($\kparb\parallel\xx$) with linear polarization in the $y-z$ plane as indicated by the angle $\phi_{\rm p}$. Each of the electron Bragg diffraction orders (see yellow arrows for one of them) splits in energy and direction of reflection/transmission upon exchange of a net number of photons $\ell$. ({\bf b})~Electron isoenergy contours after exchanging $\ell=-1$, 0, and 1 photons (blue, black, and red circles), superimposed on the reciprocal lattice of the atomic monolayer ($4\pi/\sqrt{3} d$ distance between sites). We consider 71.4~eV incident electrons and 1 eV photons. ({\bf c-h})~Intensity of transmitted (c-e) and reflected (f-h) Bragg peaks for three different incident electron energies (see labels above (c,f), (d,g), and (e,h)) upon exchange of $\ell=-1$, 0, or 1 photons (1~eV energy). The light wave vector $\kparb$ is indicated in (c) and the electric field amplitude is $E_0=2.5\times10^8$~V/m with polarization set by $\phi_{\rm p}=45\degree$. The area of the circles gives the fraction of electrons in each Bragg peak (see log-scale legend). ({\bf i-k})~Same as (e), but for varying polarization angles $\phi_{\rm p}$ (see top labels).}
\label{Fig5}
\end{figure*}

\subsection*{Light-assisted low-energy electron diffraction}

Low-energy electrons with energies $\sim10-500\,$eV are commonly used to study the atomic structure of crystal surfaces in low-energy electron diffraction~\cite{R95,P1974,P1984} (LEED) because they penetrate only a few atomic layers and have de Broglie wavelengths commensurate with the atomic spacings. In a related context, energy-resolved inelastic low-energy electron surface scattering is also used to probe surface modes~\cite{CBH92,NHH01,NYI06}, while ultrafast LEED grants one access to time-resolved structural dynamics~\cite{GSS14,VSH18}. Here, we theoretically study the interaction of surface-diffracted electrons with light plane waves and show the important role played by recoil in the underlying electron--light coupling, including the presence of lattice resonances that boost the interaction under Rayleigh anomaly conditions~\cite{R1907-2}.

To illustrate electron--light--matter interactions in the presence of Bragg diffraction, we consider a low-energy electron normally impinging on an illuminated monolayer of gold atoms arranged in a (111) triangular lattice with an Au--Au bond distance of $0.288\,$nm (Fig.~\ref{Fig5}a). In the absence of external illumination, the diffracted electron wave function consists of components with wave vectors given by $\qb^\pm_{\Gb0}=\Gb\pm\sqrt{q_0^2-G^2}\,\zz$, where $\Gb$ are 2D reciprocal lattice vectors, while the $+$ ($-$) sign corresponds to upward (downward) electron motion relative to the atomic plane. Lattice scattering is elastic, so all of these wave vectors have a magnitude $q_0=\sqrt{2\me\varepsilon_0/\hbar}$ determined by the incident electron energy $\hbar\varepsilon_0$. Diffracted electron plane waves with $G<q_0$ generate observable LEED spots, as determined by an Ewald sphere construction (see Fig.~\ref{Fig5}), whereas waves with $G>q_0$ are evanescent. The latter do not explicitly contribute to the far-field electron scattering probability, but they have to be retained in the description of dynamical electron diffraction by the atomic layer~\cite{R95,P1974,P1984}. Starting from a normally incident electron, the wave function associated with the incident and scattered waves in the absence of illumination takes the form
\begin{subequations}
\begin{align}
\psi_0(\rb,t)=\Big[\ee^{-\ii q_0z}+\sum_{\pm}B^\pm_{\Gb}\ee^{\ii\qb^\pm_{\Gb0}\cdot\rb}\Theta(\pm z)\Big]\,\ee^{-\ii\varepsilon_0t}, \label{psi0lattice}
\end{align}
where $B^\pm_{\Gb}$ are scattering amplitudes.

Upon interaction with an incident light plane wave of wave vector $\kb$, every diffraction order (either propagating or evanescent) can exchange energy with the light field in multiples of the photon energy and in-plane wave vector (i.e., $\ell\hbar\omega$, and $\ell\kparb$, respectively), giving rise to diffracted components with wave vectors $\qb^\pm_{\Gb\ell}=\Gb+\ell\kparb\pm q_{\Gb\ell z}$, where $q_{\Gb\ell z}=\sqrt{2\me(\varepsilon_0+\ell\omega)^2/\hbar-|\Gb+\ell\kparb|^2}\,\zz$, labeled by the direction of motion ($+/-$ for upward/downward scattering) and the net lattice and photon momentum exchanges ($\hbar\Gb$ and $\ell\hbar\kparb$). The total electron wave function takes the form
\begin{align}
\psi(\rb,t)=\Big[\ee^{-\ii q_0z}+\sum_{\pm}C^\pm_{\Gb\ell}\ee^{\ii\qb^\pm_{\Gb\ell}\cdot\rb-\ii\ell\omega t}\Theta(\pm z)\Big]\,\ee^{-\ii\varepsilon_0t}, \label{psilattice}
\end{align}
\end{subequations}
where the amplitudes $C_{\Gb\ell}^\pm$ are self-consistently determined from an extension of LEED theory to incorporate the interaction with both the atomic lattice and the optical field [see SI for details, including analytical expressions for the coefficients $B^\pm_{\Gb}$ and $C^\pm_{\Gb\ell}$ in Eqs.~(\ref{psi0lattice}) and (\ref{psilattice}), respectively.] For simplicity, we limit our analysis to first order in the electron--light interaction, but we incorporate the interaction with the lattice to all orders.

As the electron energy increases, the number of diffracted spots also increases because more points are inside the Ewald sphere. In the elastic part ($\ell=0$), a given diffraction order $\Gb$ is observed in the LEED pattern when the electron energy exceeds a threshold energy $\hbar^2G^2/2\me$. However, in the inelastic components, the changes in electron energy and momentum enter the condition for far-field propagation. For example, after exchanging $\ell$ photons, a previously evanescent order may become propagating if $\hbar|\Gb+\ell\kparb|^2/2\me<\varepsilon_0+\ell\omega$, and likewise, a propagating Bragg-diffracted beam may become evanescent (Fig~\ref{Fig5}b). These effects are more relevant when $\varepsilon_0\sim\hbar G^2/2\me$. In addition, electrons scattered at the onset of a diffraction order move under grazing conditions, so they spend more time near the surface and, therefore, undergo a stronger interaction with light. Consequently, by choosing the electron energy close to the threshold of one of the $\Gb$ beams, we expect to increase the coupling of diffracted electrons to light, emphasizing the importance of recoil (see below).

This phenomenology is illustrated in the calculations presented in Fig.~\ref{Fig5}c-h for three different incident electron energies (in separate columns) near the threshold of the $(1,1)$ and its symmetry-equivalent diffraction spots. We show the intensities of transmitted (Fig.~\ref{Fig5}c-e) and reflected (Fig.~\ref{Fig5}f-h) beams. Following the conditions under which a maximum electron--light interaction was observed in Fig.~\ref{Fig4}, we take the electron to be normally impinging on the atomic plane, while a light plane wave is incident parallel to the $x$ surface direction with linear polarization as indicated in Fig.~\ref{Fig5}a. The atomic lattice is oriented with Au--Au bonds along $y$. Upon inspection of our numerical results, we estimate an overall inelastic scattering probability of $\sim10-20\%$ for incident electron energies up to 50~eV and optical electric-field strengths $E_0\sim2.5\times10^8\,$V/m. Additionally, we find that the waves corresponding to $\ell=\pm1$ typically have comparable intensities to the transmitted $\ell=0$ beam. In contrast to the amplitude found when adopting the nonrecoil approximation, which only depends on the amplitude of each $\Gb$-dependent LEED spot and its multiplexing into different energy sidebands according to the corresponding electron--light coupling coefficient $\beta$ (i.e., considering the time integral of the field along the classical incoming and Bragg-reflected electron paths), a full quantum treatment including recoil reveals that all diffraction orders (propagating and evanescent) can contribute.

\begin{figure}[ht!]
\centering\includegraphics[width=0.45\textwidth]{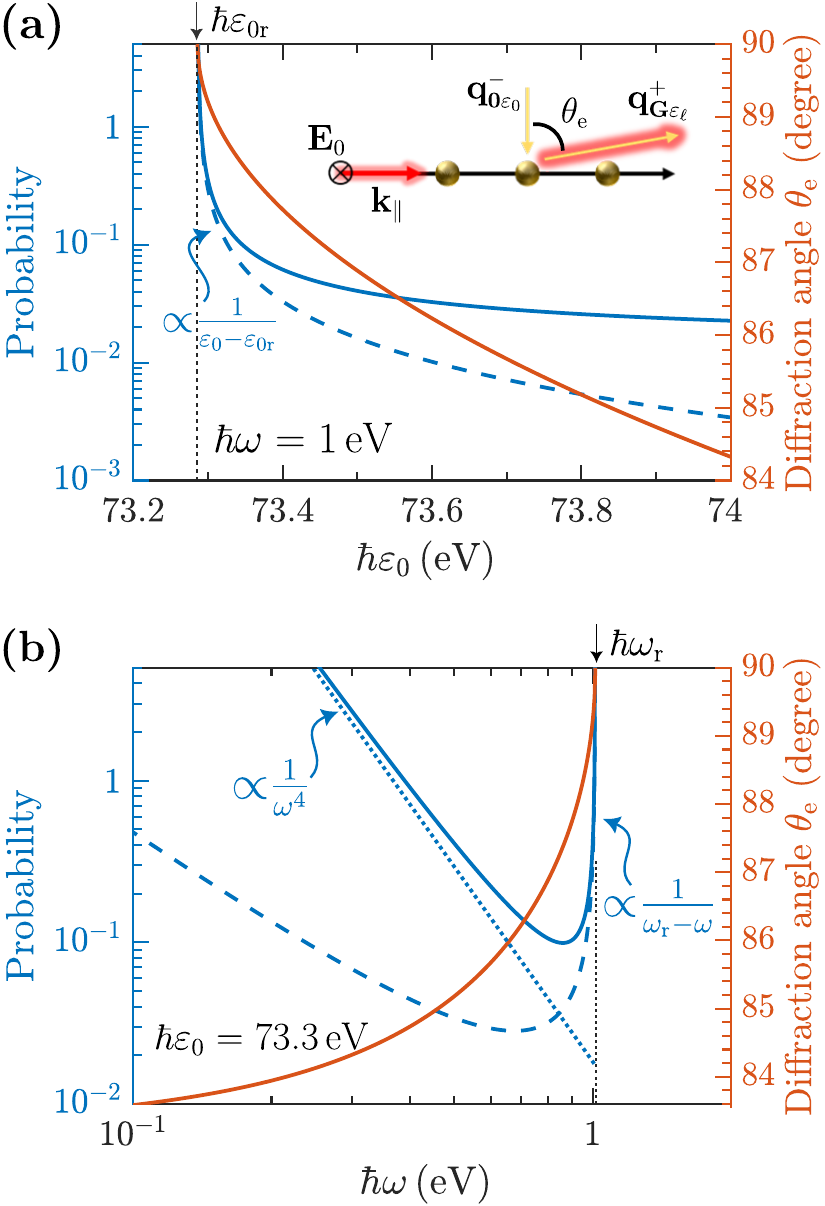}
\caption{\textbf{Strong coupling between light and diffracted electrons.} ({\bf a})~Probability (blue curve, to first order) and azimuthal outgoing angle (orange curves) of the $(2,1)$ diffraction spot [$\Gb=(2\pi/d)\,\big(\sqrt{3}\xx-\yy\big)$] after emitting one photon ($\ell=-1$) as a function of electron energy $\hbar\varepsilon_0$ under the configuration of Fig.~\ref{Fig5}a for a light plane wave impinging along $x$ with polarization along $y$, amplitude $E_0=2.5\times 10^7$~V/m, and photon energy $\hbar\omega=1$~eV. Rigorous first-order theory (solid-blue curve) is compared with the analytical approximation in Eq.~(\ref{Canalytical}) (dashed curve). ({\bf b})~Same as (a), but shown as a function of photon energy for an incident electron energy $\hbar\varepsilon_0=73.3$~eV. A divergent probability is observed when the inelastically diffracted beam becomes grazing at electron and photon energies $\hbar\varepsilon_{0{\rm r}}$ and $\hbar\omega_{\rm r}$ in (a) and (b), respectively. The probability also diverges at low frequencies with the asymptotic behavior $\propto1/\omega^4$ indicated by the dotted-blue curve in (b).}
\label{Fig6}
\end{figure}

The onset of a new diffraction order during the scattering of waves by a periodic structure causes an anomaly consisting of the depletion of the specularly reflected and directly transmitted beams, as pointed out by Lord Rayleigh in the context of light diffraction by periodic gratings~\cite{R1907-2,R1907,paper090}. In light-assisted inelastic electron diffraction, a related anomaly takes place when a scattered beam becomes grazing (i.e., a vanishing out-of-plane wave vector component $q_{\Gb\varepsilon_\ell z}=0$ for a combination of reciprocal lattice vector $\Gb$ and sideband order $\ell$, or equivalently, $\hbar\varepsilon_0+\ell\hbar\omega=(\hbar^2/2\me)|\Gb+\ell\kparb|^2$ under normal electron-incidence conditions). Upon examination of the corresponding coefficients $C_{\Gb\ell}^\pm$ in Eq.~(\ref{psilattice}) (see SI), considering $q_{\Gb\varepsilon_\ell z}\approx0$ and taking the field amplitude $\Eb_0\perp\zz$ for simplicity, we can approximate
\begin{align}
C_{\Gb,\ell}^\pm\approx&\frac{\ii^\ell e}{\hbar\omega} \;
\frac{\Eb_0\cdot\Gb}{q_{\Gb\varepsilon_\ell z}\;q_{\Gb\varepsilon_0z}} \Big(B^+_{\Gb}+B^-_{\Gb}\Big),
\label{Canalytical}
\end{align}
and consequently, the probability $\big|C_{\Gb,\ell}^\pm\big|^2$ diverges as $1/q_{\Gb\varepsilon_\ell z}^2$. An illustrative example is presented in Fig.~\ref{Fig6} when varying either the electron energy (Fig.~\ref{Fig6}a) or the photon energy (Fig.~\ref{Fig6}b) around the conditions for the grazing emission of an inelastic electron beam. A divergence in the probability calculated to the first order of interaction with the light is observed, leading to unphysical values above unity. This indicates that the system enters into the nonpertubative regime, requiring higher orders of interaction to describe the system, but also revealing that strong electron--light coupling can be reached even for small light intensities. Such a strong interaction results from in-phase scattering by a large number of atoms in the planar lattice, which demands the use of sufficiently broad electron and light beams that can be regarded as plane waves over a wide surface area. This divergence is well captured by Eq.~(\ref{Canalytical}) [Fig.~\ref{Fig6}, dashed-blue curves], in reasonable agreement with our rigorous first-order results [Fig.~\ref{Fig6}, solid-blue curves, obtained from Eq.~(\ref{S23}) in Appendix]. We note that, in addition to the explicit $1/\omega$ factor in Eq.~(\ref{Canalytical}), the inelastic scattering coefficient $C_{\Gb,\ell}^\pm$ is dominated by a $1/(q_{\Gb\varepsilon_l z}-q_{\Gb\varepsilon_0 z})$ term in Eq.~(\ref{S23}) (see SI), thus resulting in an overall $\propto1/\omega^4$ scaling of the probability with decreasing photon frequency $\omega$, as indicated in Fig.~\ref{Fig6}b.

\section*{Discussion}

In summary, based on a comprehensive theoretical treatment of the quantum-mechanical interaction between low-energy free electrons and optical fields with comparable photon energies, we have identified a plethora of recoil and quantum effects emerging in the form of dramatic modifications in the energy and angular distribution of electrons undergoing elastic surface scattering and inelastic interaction with optical fields associated with either surface polaritons or propagating light. In particular, we have shown that free electrons interacting with evanescent optical fields can undergo classically forbidden backscattering. Furthermore, the interaction between surface-scattered electrons and unscattered light plane waves renders a nonzero electron--light coupling due to the breaking of translational symmetry in the electron wave function; we propose that suspended atomically thin layers may provide suitable conditions (high transparency to light and large electron scattering) to observe this effect.

As a common element in the interaction between free electrons and crystal surfaces, we have incorporated Bragg diffraction, which leads to an interplay between scattering by the atomic lattice and inelastic photon exchanges. The latter can transform propagating diffraction orders into evanescent or the other way around. Importantly, strong electron--light coupling is predicted at the onset of an inelastically scattering electron beam, capitalizing on the in-phase interaction with many atoms in the structure. Such a strong coupling regime could be leveraged to optically shape free electrons using moderate light intensities, potentially operating in the continuous-wave regime without damaging the scattering material.

Although we have focused on planar surfaces, the concept of combining elastic scattering of low-energy electrons by a material structure and the inelastic interaction with light of comparably low photon energy is more general and could involve the use of non-periodic nanostructures such as holes, tips, and other curved elements to guide and reshape the electron wave function and also increase or spatially modulate its interaction with specific optical modes. In a related context, electron interaction with illuminated atoms in the gas phase has a long tradition \cite{WHC1977,WHS1983,FJ1990,paper387}, which could be revisited as a platform to optically modulate electrons and explore new physics. The novel directions opened by the presented theory and simulations could be experimentally explored in currently available low-energy electron-microscope setups.

\section*{Acknowledgments}
The authors acknowledge insightful discussions with V. Di Giulio, S. V. Yalunin, and J. Otto. This work has been supported in part by the European Research Council (Advanced Grants 789104-eNANO and 101055435-ULEEM), the European Commission (Horizon 2020 Grants No. 101017720 FET-Proactive EBEAM and No. 964591-SMART-electron), the Spanish MICINN (PID2020-112625GB-I00 and Severo Ochoa CEX2019-000910-S), the Catalan AGAUR (Grant No. 2023 FI-1 00052) and CERCA Programs, and Fundaci\'{o}s Cellex and Mir-Puig.

\appendix
\begin{widetext}

\section{Low-energy electron scattering by an illuminated homogeneous surface}
\renewcommand{\thefigure}{A\arabic{figure}} 
\renewcommand{\theequation}{A\arabic{equation}} 
\renewcommand{\thetable}{A\arabic{table}} 
\renewcommand{\thesection}{A\arabic{section}} 
\label{S1}

We present a self-contained theory of low-energy electron scattering at planar interfaces that are either supporting surface polaritons or subject to external illumination.

\subsection{Electron scattering by a planar surface}

We consider an electron of energy $\hbar \varepsilon_0$ elastically scattered by a homogeneous planar structure represented through a one-dimensional (1D) potential energy $V(z)$. Different types of surfaces are discussed below, with the electron impinging from the vacuum region ($z>0$) as a plane wave of well-defined in-plane wave vector $\q0parb$ and total wave vector $q_0=\sqrt{2\me \varepsilon_0/\hbar}$. The incident out-of plane wave vector is then $-q_{0z}$ with $q_{0z}=\sqrt{q_0^2-q_{0\parallel}^2}$. The Hamiltonian of the system can be written as
\begin{align}
\Hop_0(\rb)=-\frac{\hbar^2\nabla^2}{2\me}+V(z), \nonumber
\end{align}
with $V(z)=0$ in the $z>0$ region. Translational invariance in the $x-y$ plane allows us to factorize the electron wave function as
\begin{align}
\psi_0(\rb,t)=\ee^{\ii\q0parb\cdot\Rb-\ii\varepsilon_0t}\,\psi_0(z), \label{psidecomposition}
\end{align}
where $\Rb=(x,y)$ and the subscript in $\psi_0$ indicates that the interaction with light is still not included.\\

In the formalism that follows, $\Hop_0(\rb)$ determines $\psi_0(z)$ starting from an incident electron plane wave, as well as the noninteracting Green function $\mathcal{G}_0(\rb,\rb',t-t')$ implicitly defined by~\cite{E06}
\begin{align}
\big[\hat{\mathcal{H}}_0(\rb)-\ii\hbar\partial_t\big]\,\mathcal{G}_0(\rb,\rb',t-t')=-\delta(\rb-\rb')\delta(t-t'). \label{HtGd}
\end{align}
Time invariance allows us to write the Green function in frequency space as $\mathcal{G}_0(\rb,\rb',t-t')=(2\pi)^{-1}\int\dd\varepsilon\;\ee^{\ii\varepsilon(t'-t)}\,\mathcal{G}_0(\rb,\rb',\varepsilon)$. Likewise, surface translation symmetry reduces the dependence on in-plane coordinates to the difference $\Rb-\Rb'$. Moving to Fourier space, this dependence can be expressed as a combination of $\ee^{\ii\qparb\cdot(\Rb-\Rb')}$ waves with in-plane wave vectors $\qparb$. In addition, the Hamiltonian $\Hop_0(\rb)$ depends on $\Rb$ only through the $\nabla^2_\Rb$ term, so we can separate the frequency in parallel and perpendicular components as $\varepsilon=\hbar q_\parallel^2/2\me+\varepsilon^\perp$. Putting these elements together, we can write
\begin{align}
\mathcal{G}_0(\rb,\rb',t-t')=\int\frac{\dd^2\qparb}{(2\pi)^2}\; \ee^{\ii\qparb\cdot(\Rb-\Rb')}\int\frac{\dd\varepsilon}{2\pi}\; \ee^{\ii\varepsilon(t'-t)}
\;\mathcal{G}_0\big(z,z',\varepsilon-\hbar q_\parallel^2/2\me\big), \label{eq:G3D_to_G1D}
\end{align}
which, upon insertion into Eq.~(\ref{HtGd}), leads to the equation
\begin{align}
\bigg[-\frac{\hbar^2}{2\me}\partial_z^2+V(z)-\hbar\varepsilon^\perp\bigg]\,\mathcal{G}_0(z,z',\varepsilon^\perp)=-\delta(z-z') \label{Green1Deq}
\end{align}
for the 1D Green function $\mathcal{G}_0(z,z',\varepsilon^\perp)$.\\

\begin{subequations}
\label{Greenfunctions}
We consider three different types of potentials, for which the Green function can be calculated analytically as follows:
\begin{itemize}
\item {\it Finite potential barrier.} To describe atomically thin two-dimensional (2D) materials, we set $V(z)=U_0\delta(z)$. The parameter $U_0$ has units of energy times length. Arguing that an atomic monolayer can be described by a barrier of finite thickness $d\lesssim1$~nm and internal potential $V_0$ in the eV range, we expect $U_0\approx V_0d$ in the eV$\times$nm range when adopting the zero-thickness approximation. The corresponding Green function reads
\begin{align}
\mathcal{G}_0(z,z',\varepsilon^\perp)&=-\frac{\ii\me}{\hbar^2q_z}\,\Big[\ee^{\ii q_z|z-z'|}+r_{q_z}\,\ee^{\ii q_z(|z|+|z'|)}\Big], \label{Gdelta}
\end{align}
where $q_z=\sqrt{2\me\varepsilon^\perp/\hbar}$ is the out-of-plane wave vector, the first term accounts for free electron propagation, and the second term is proportional to the energy-dependent reflection coefficient of the potential barrier $r_{q_z}=\big(\ii\hbar^2|q_z|/\me U_0-1\big)^{-1}$. Upon inspection, Eq.~(\ref{Gdelta}) can be readily verified to satisfy Eq.~(\ref{Green1Deq}) for the chosen potential.
\item {\it Infinite potential barrier.} As an instructive limit, we also consider complete surface electron reflection, which can be realized through the potential barrier introduced above by setting $U_0=\infty$ (i.e., with $r_{q_z}=-1$). From Eq.~(\ref{Gdelta}), the Green function now reduces to
\begin{align}
\mathcal{G}_0(z,z',\varepsilon^\perp)&=-\frac{\ii\me}{\hbar^2q_z}\big(\ee^{\ii q_z|z-z'|}-\,\ee^{\ii q_z(|z|+|z'|)}\big) \label{Ginfinite} \\
&=-\frac{\ii\me}{\hbar^2q_z}\left(\ee^{\ii q_z|z-z'|}-\,\ee^{\ii q_z(|z|+|z'|)}\right)\,\big[\Theta(z)\,\Theta(z')+\Theta(-z)\,\Theta(-z')\big], \nonumber
\end{align}
where the second line shows that positive and negative $z$ regions do not mix [i.e., an electron coming from $z>0$ stays fully inside that region after scattering, as one can verify from Eq.~(\ref{finalrecusion}) below]. 
\item {\it Free electron.} When the interaction with the material can be neglected, we set $V(z)=0$ (i.e., $U_0=0$), so we have $r_{q_z}=0$, and therefore, Eq.~(\ref{Gdelta}) reduces to
\begin{align}
\mathcal{G}_0(z,z',\varepsilon^\perp)&=-\frac{\ii \me}{\hbar^2q_{z}}\ee^{\ii q_{z}|z-z'|}. \label{Gfree}
\end{align}
\end{itemize}
\end{subequations}
In addition, the initial wave function $\psi_0(z)$ produced by an incident wave $\ee^{-\ii q_{0z}}$ also depends on the type of potential and takes the explicit form
\begin{align}
\psi_0(z)=\left\{\begin{array}{ll}
\big[\ee^{-\ii q_{0z}z}+r_{q_{0z}}\,\ee^{\ii q_{0z}z}\big]\,\Theta(z)+t_{q_{0z}}\,\ee^{-\ii q_{0z}z}\,\Theta(-z),
   &\quad\quad\quad\text{(finite potential barrier)} \\
\big[\ee^{-\ii q_{0z}}-\ee^{\ii q_{0z}}\big]\,\Theta(z),
   &\quad\quad\quad\text{(infinite potential barrier)} \\
\ee^{-\ii q_{0z}},
   &\quad\quad\quad\text{(free electron)}
\end{array}\right. \label{psi0}
\end{align}
where $t_{q_z}=1+r_{q_z}$ is the transmission coefficient (evaluated at the incident out-of-plane wave vector $q_z=-q_{0z}$), which guarantees the continuity of $\psi_0(z)$. In the analysis presented below, we use the 1D Green function and initial wave function for the finite potential barrier, from which the free-electron and infinite-potential-barrier configurations are directly obtained by setting $r_{q_z}=0$ and $r_{q_z}=-1$, respectively.\\

\subsection{Interaction of light}

The optical field is introduced through a classical vector potential $\Ab(\rb,t)=\Ab(z)\,\ee^{\ii\kparb\cdot\Rb-\ii\omega t}+{\rm c.c.}$ with well-defined in-plane wave vector $\kparb$ and frequency $\omega$. By working in a gauge with vanishing scalar potential, adopting the minimal-coupling prescription, and neglecting $A^2$ terms, the interaction Hamiltonian reduces to
\begin{align}
\Hop_1(\rb,t)=\frac{-\ii e\hbar}{\me c}\Ab(\rb,t)\cdot\bm{\nabla}, \label{H1}
\end{align}
so the total Hamiltonian becomes $\Hop(\rb,t)=\Hop_0(\rb)+\Hop_1(\rb,t)$. The electron wave function $\psi(\rb,t)$ is then modified relative to $\psi_0(\rb,t)$ as described by the Lippmann--Schwinger (LS) equation~\cite{M1966,S94} $\psi(\rb,t)=\psi_0(\rb,t) + \int\dd^3\rb'\dd t'\;\mathcal{G}_0(\rb,\rb',t-t')\,\Hop_1(\rb',t')\,\psi(\rb',t')$. At this point, it is convenient to write the perturbation series
\begin{align}
\psi(\rb,t) = \sum_{n=0}^{\infty} \psi^{(n)}(\rb,t), \label{eq:def_wf_n}
\end{align}
where the $n$-th term is order $n$ in the vector potential $\Ab$. By inserting Eq.~(\ref{eq:def_wf_n}) into the LS equation, we find the recursion formula
\begin{align}
    \psi^{(n)}(\rb,t)= \int \dd^3\rb' \int \dd t' \, \mathcal{G}_0(\rb,\rb',t-t')\,\Hop_1(\rb',t') \, \psi^{(n-1)}(\rb',t') \label{eq:LippSchwEq}
\end{align}
for $n>0$, starting with $\psi^{(0)}\equiv\psi_0$.\\

Because the vector potential depends on time through $\ee^{\pm \ii \omega t}$, each subsequent perturbation order introduces the possibility of absorbing or emitting one additional photon. Hence, at a given order $n$ the electron energy can be modified to $\hbar(\varepsilon_0+\ell\omega)$ with $|\ell|\le n$. Using this and further imposing conservation of in-plane momentum, we can decompose the post-interaction wave function as 
\begin{align}
\psi(\rb,t) = \sum_{n=0}^\infty\sum_{\ell=-n}^n \varphi_{n\ell}(z) \, \ee^{ \ii (\q0parb + \ell \,\kparb) \cdot \Rb}\, \ee^{-\ii (\varepsilon_0 + \ell \omega) t}. \label{eq:def_wf_n_l}
\end{align}
We now introduce Eqs.~\eqref{eq:G3D_to_G1D} and \eqref{eq:def_wf_n_l} into Eq.~\eqref{eq:LippSchwEq}, use the factorization in Eq.~(\ref{psidecomposition}), and identify terms with the same $(\Rb,t)$ dependence on both sides of the equation. This procedure shows that Eq.~(\ref{eq:def_wf_n_l}) is indeed a solution when the $z$-dependent coefficients are determined by iterating the relation
\begin{align}
\varphi_{n\ell}(z)=\frac{\hbar e}{\me c} \int \dd{z'}\, \mathcal{G}_0(z,z',\varepsilon^\perp_\ell) \; \Big\{
\Ab(z') &\cdot \big[\q0parb + (\ell-1)\,\kparb -\ii \zz \, \partial_{z'} \big] \,\varphi_{n-1,\ell-1}(z') \nonumber\\
+\Ab^*(z') &\cdot \big[\q0parb + (\ell+1)\,\kparb -\ii \zz \, \partial_{z'} \big] \,\varphi_{n-1,\ell+1}(z')
\Big\}, \label{finalrecusion} 
\end{align}
where $\varepsilon_\ell^\perp = \varepsilon_0 + \ell \omega - \hbar |\qparb + \ell\, \kparb|^2/2\me$. Equation~(\ref{finalrecusion}), which is reproduced as Eq.~(2) in the main text, applies to $n>0$ with a seeding term $\varphi_{00}(z)=\psi_0(z)$ and imposing $\varphi_{n\ell}(z)=0$ for $|\ell|>n$.\\

For simplicity, we assume the in-plane wave vectors of light and electron to be both along the surface direction $x$ (i.e., $\q0parb=q_{0\parallel}\xx$ and $\kparb=\kpar\xx$). In what follows, we consider two types of optical fields:
\begin{subequations}
\label{vectorfields}
\begin{itemize}
\item {\it Unscattered light plane waves.} For sufficiently thin films (e.g., self-standing atomic monolayers), light scattering by the material can be approximately neglected. Then, taking an incident wave vector $\kb=\kpar\xx-k_z\zz$ and an electric field amplitude $\Eb_0$, the $z$ dependence of the vector potential reduces to
\begin{align}
\Ab(z)=\frac{1}{\ii k}\Eb_0\,\ee^{\ii k_zz}.
\end{align}
We consider propagating light with $\kpar\le k=\omega/c$ and $k_z=\sqrt{k^2-\kpar^2}$.
\item {\it Surface polaritons.} For films of negligible thickness, the electric field associated with a polariton of in-plane wave vector $\kparb=\kpar\xx$ and frequency $\omega$ can be written as $\Eb(\rb,t)=\Eb(\rb)\,\ee^{-\ii\omega t}+{\rm c.c.}$ with $\Eb(\rb)=E_0\;\big(\kpar^2+\kappa^2\big)^{-1/2}\,\big(\ii\,\kappa\,\xx-\sign\{z\}\,\kpar\,\zz\,\big)\,\ee^{\ii\kpar x-\kappa|z|}$, where $E_0$ is a global amplitude and $\kappa=\sqrt{\kpar^2-k^2}$ determines an out-of-plane exponential decay with $\kpar>k$. This expression is constructed to satisfy $\nabla\cdot\Eb(\rb,t)=0$ at $z\neq0$. Then, the vector potential reduces to $\Ab(\rb)=\Eb(\rb)/\ii k=\Ab(z)\,\ee^{\ii\kpar x}$, where the $z$ dependence is given by
\begin{align}
\Ab(z)=\frac{E_0}{k}\frac{1}{\sqrt{\kappa^2+\kpar^2}}\;\Big(\kappa\xx+\ii\,\kpar\sign\{z\}\,\zz\Big)\,\ee^{-\kappa|z|}.
\end{align}
We find it convenient to express the in-plane wave vector as $\kpar=\neff k$ in terms of an effective refractive index $\neff$. In this work, we take $\neff\gg1$ values, so have $\kappa\approx\kpar$, and therefore, $\Ab(z)\approx\big(E_0/\sqrt{2}\,k\big)\;\big(\xx+\ii\,\sign\{z\}\,\zz\big)\,\ee^{-\kpar|z|}$. However, we use the fully retarded expressions in our calculations. Incidentally, these expressions can also be used for arbitrary planar material systems contained within the $z<0$ region if we are only interested in the outer $z>0$ half-space (i.e., without electron penetration inside the material).
\end{itemize}
\end{subequations}

Next, we solve Eq.~(\ref{finalrecusion}) by plugging the expressions for $\mathcal{G}_0(z,z',\varepsilon^\perp_\ell)$ and $\Ab(z)$ given in Eqs.~(\ref{Greenfunctions}) and (\ref{vectorfields}).

\subsection{Recursive solution for the self-consistent electron wave function}
\label{rrecursive}

We solve Eq.~(\ref{finalrecusion}) by expanding the wave function components $\varphi_{n\ell}(z)$ following a recursive procedure to determine their explicit dependence on $z$, the parameters involved in such dependence, and how they are modified at each iteration step given the analytical expressions for $\mathcal{G}_0(z,z',\varepsilon^\perp_\ell)$ and $\Ab(z)$ in Eqs.~(\ref{Greenfunctions}) and (\ref{vectorfields}). Because these quantities depend on $z$ through exponential factors, and therefore, the $z'$ integral at each iteration $n$ in Eq.~(\ref{finalrecusion}) produces additional exponential factors consisting of a combination of those at order $n-1$ plus those of $\mathcal{G}_0(z,z',\varepsilon^\perp_\ell)$ and $\Ab(z)$, the wave function components must take the general form
\begin{align}
\varphi_{n\ell}(z)=\sum_{s=\pm1}\sum_{j=1}^{N_{n\ell}} \alpha_{n \ell}^{js} \, \ee^{\zeta_{n\ell}^{js}z}\,\Theta{(sz)} \label{phinlexpansion}
\end{align}
[Eq.~(3) in the main text], where the number of terms in this sum ($N_{n\ell}$) is obviously increasing with $n$ and also depends on $\ell$. Specifically, the vector potentials under consideration can be written as
\begin{align}
\Ab(z)=\sum_{s=\pm1}\Ab^s \, \ee^{\eta^s z}\,\Theta{(sz)}, \label{Aexpansion}
\end{align}
where the coefficients $\Ab^{\pm1}$ and $\eta^{\pm1}$ can be directly identified from Eqs.~(\ref{vectorfields}). Likewise, the Green function $\mathcal{G}_0(z,z',\varepsilon^\perp_\ell)$ is given by Eq.~(\ref{Gdelta}) with the reflection coefficient $r_{q_{\ell z}}$ set to different values depending on the choice for the potential $V(z)$. Note that in the $\ell$ channel, the out-of-plane wave vector is given by $q_{\ell z}=\sqrt{2\me\varepsilon^\perp_\ell/\hbar+\ii0^+}=\sqrt{2\me(\varepsilon_0+\ell\omega)/\hbar-|\qparb + \ell\, \kparb|^2+\ii0^+}$, where the square root is taken to yield a positive imaginary part. By inserting Eqs.~(\ref{Gdelta}), (\ref{phinlexpansion}), and (\ref{Aexpansion}) into Eq.~(\ref{finalrecusion}), we obtain the equivalent recursion formula (for $\pm z>0$)
\begin{align}
\sum_j \alpha_{n \ell}^{j\pm} \, \boxed{\ee^{\zeta_{n\ell}^{j\pm}z}}=
&-\frac{\ii e}{\hbar cq_{\ell z}}
\sum_{s=\pm1} \sum_j
\Bigg\{
2\ii q_{\ell z}\delta_{s,\pm1}
\Bigg[\boxed{\ee^{(\eta^s+\zeta_{n-1,\ell-1}^{js})z}}\;\Ab^s\cdot\big[\q0parb + (\ell-1)\,\kparb -\ii \zz \, \zeta_{n-1,\ell-1}^{js} \big] \,
\frac{\alpha_{n-1,\ell-1}^{js}}{q_{\ell z}^2+\big(\eta^s+\zeta_{n-1,\ell-1}^{js}\big)^2} \nonumber\\
&\quad\quad\quad+\boxed{\ee^{(\eta^{s*}+\zeta_{n-1,\ell+1}^{js})z}}\;\Ab^{s*}\cdot\big[\q0parb + (\ell+1)\,\kparb -\ii \zz \, \zeta_{n-1,\ell+1}^{js} \big] \,
\frac{\alpha_{n-1,\ell+1}^{js}}{q_{\ell z}^2+\big(\eta^{s*}+\zeta_{n-1,\ell+1}^{js}\big)^2} \Bigg] \label{finaliteration}\\
+\boxed{\ee^{\pm\ii q_{\ell z}z}} \;
\Bigg[&\Ab^s\cdot\big[\q0parb + (\ell-1)\,\kparb -\ii \zz \, \zeta_{n-1,\ell-1}^{js} \big] \,\alpha_{n-1,\ell-1}^{js}\;
\Bigg(\frac{-r_{q_{\ell z}}}{\ii q_{\ell z}+s\;\big(\eta^s+\zeta_{n-1,\ell-1}^{js}\big)}
\pm\frac{s}{\ii q_{\ell z}\mp\big(\eta^s+\zeta_{n-1,\ell-1}^{js}\big)}\Bigg) \nonumber\\
+&\Ab^{s*}\cdot\big[\q0parb + (\ell+1)\,\kparb -\ii \zz \, \zeta_{n-1,\ell+1}^{js} \big] \,\alpha_{n-1,\ell+1}^{js}\;
\Bigg(\frac{-r_{q_{\ell z}}}{\ii q_{\ell z}+s\;\big(\eta^{s*}+\zeta_{n-1,\ell+1}^{js}\big)}
\pm\frac{s}{\ii q_{\ell z}\mp\big(\eta^{s*}+\zeta_{n-1,\ell+1}^{js}\big)}\Bigg) \Bigg] \Bigg\}, \nonumber 
\end{align}
where we have performed the $z'$ integrals using the identities
\begin{subequations}
\label{integrals}
\begin{align}
&\int\dd{z'}\,\Theta{(\pm z')}\,\ee^{\ii q_{\ell z}|z'|+\Delta z'}=\frac{-1}{\ii q_{\ell z}\pm\Delta}, \\
&\int\dd{z'}\,\Theta{(\pm z')}\,\ee^{\ii q_{\ell z}|z-z'|+\Delta z'}=\frac{\pm\sign\{z\}}{\ii q_{\ell z}-\sign\{z\}\Delta}\,\ee^{\ii q_{\ell z}|z|} + \frac{2\ii q_{\ell z}}{q_{\ell z}^2+\Delta^2}\,\ee^{\Delta z}\;\Theta(\pm z).
\end{align}
\end{subequations}
Incidentally, the exponentials inside the integrals vanish at $|z'|\to\infty$ because they contain either evanescent electron/light components or propagating fields in which the electron wave vectors $q_{\ell z}$ possess an arbitrarily small positive imaginary part that makes them consistent with the use of retarded Green functions. At every iteration step $n>0$, we solve Eq.~(\ref{finaliteration}) by setting the coefficients $\alpha_{n \ell}^{j\pm}$ and $\zeta_{n\ell}^{j\pm}$ in the left-hand side to match the exponential terms in the right-hand side (see boxed expressions), starting with the $n=0$ seeding values $\big\{\alpha_{00}^{1,+1}=1,\,\zeta_{00}^{1,+1}=-\ii q_{0z}\big\}$, $\big\{\alpha_{00}^{2,+1}=r_{q_{0z}},\,\zeta_{00}^{2,+1}=\ii q_{0z}\big\}$, and $\big\{\alpha_{00}^{1,-1}=t_{q_{0z}},\,\zeta_{00}^{1,-1}=-\ii q_{0z}\big\}$, as determined from Eq.~(\ref{psi0}), and noticing that $|\ell|\le n$.

When considering polaritons, for which $\eta^s$ in Eq.~(\ref{Aexpansion}) has a finite real part that makes the optical field evanescent, upon inspection of the numerical solution of Eq.~(\ref{finaliteration}), we corroborate that the obtained exponential coefficients of all propagating components in Eq.~(\ref{finaliteration}) (i.e., with vanishing ${\rm Re}\{\zeta_{n\ell}^{js}\}$) satisfy $\zeta_{n\ell}^{js}=\ii\,sq_{\ell z}$, as expected from energy conservation after a net exchange of $\ell$ photons. The remaining coefficients are found to possess finite real parts satisfying the condition $s\,{\rm Re}\{\zeta_{n\ell}^{js}\}<0$ (i.e., they are evanescent).

However, under external illumination, Eq.~(\ref{Aexpansion}) comprises propagating components (i.e., imaginary coefficients $\eta^s$), so the interaction region extends indefinitely in $z$, affecting the electron in the way described by Volkov \cite{W1935_2}. A direct solution of Eq.~(\ref{finaliteration}) is then problematic because one needs to separate surface-mediated inelastic transitions from far-field electron--light interactions. In actual experiments, the range of the applied light fields is limited to a finite region near the surface (e.g., when using a laser beam), so we find it more convenient to introduce a finite real part in the exponential coefficients as ${\rm Re}\{\eta^s\}=-s\gamma$ with $\gamma>0$ to prevent electron exposure to far optical fields. Then, we iterate Eq.~(\ref{finaliteration}) to the desired order $n$, perform the $|z|\to\infty$ limit to calculate the scattered electron intensity, and finally take $\gamma\to0^+$. In practice, this amounts to calculating the electron intensity after eliminating the terms of Eq.~(\ref{Aexpansion}) in which $\zeta_{n\ell}^{js}$ depends explicitly on $\gamma$. The remaining far-field terms also satisfy $\zeta_{n\ell}^{js}=\ii\,sq_{\ell z}$.

\section{Low-energy electron scattering by an illuminated atomic monolayer including diffraction}
\renewcommand{\thefigure}{B\arabic{figure}} 
\renewcommand{\theequation}{B\arabic{equation}} 
\renewcommand{\thetable}{B\arabic{table}} 
\renewcommand{\thesection}{B\arabic{section}} 
\label{S2}

While in Sec.~\ref{S1} we discussed homogenous surfaces, the in-plane atomic corrugation needs to be incorporated in a more realistic scenario. Next, we extend the theory to incorporate diffraction by an illuminated monoatomic layer.

\subsection{Low-energy electron diffraction by an atomic monolayer}

Low-energy electron diffraction (LEED) is a well-established technique for the determination of surface atomic structures based on the measurement and comparison with the theory of the electron-energy dependence of the intensities associated with different Bragg reflections by periodic atomic lattices \cite{P1974,VWC1986}. Here, we borrow the theoretical methods developed to simulate LEED and describe electron diffraction by an atomic monolayer~\cite{P1974}. Following the notation of Sec.~\ref{S1}, we consider an electron beam of energy $\hbar\varepsilon_0$ impinging on a crystalline atomic monolayer with wave vector $\qb_0=\q0parb-q_{0z}\,\zz$, where $q_{0z}=\sqrt{q_0^2-q_{0\parallel}^2}$. For simplicity, we assume a simple lattice of the same type of atoms, periodically arranged at positions $\Rb_\alpha$ in the $z=0$ plane and with unit cell area $A$. The Hamiltonian of the system can be approximated as
\begin{align}
\Hop_0(\rb)=-\frac{\hbar^2\nabla^2}{2\me}+\sum_\alpha V_{\rm atom}(|\rb-\Rb_\alpha|), \nonumber
\end{align}
where $V_{\rm atom}(r)$ is the atomic potential, considered to be isotropic. The electron is diffracted by the lattice and exchanges discrete amounts of in-plane momentum $\hbar\Gb$ determined by the 2D reciprocal lattice vectors $\Gb$, resulting in scattered beams with three-dimensional wave vectors $\q0parb+\Gb\pm q_{\Gb\varepsilon_0 z}\zz$ directed along upward ($+$) and downward ($-$) directions. The out-of-plane wave vector component $q_{\Gb\varepsilon_0 z}=\sqrt{q_0^2-|\q0parb+\Gb|^2+\ii0^+}$ is determined by energy conservation with the square root taken to yield a positive imaginary part. For $|\q0parb+\Gb|<q_0$, the diffracted waves propagate to the far field, giving rise to observable LEED spots, whereas waves with $|\q0parb+\Gb|>q_0$ are evanescent. The total electron wave function outside the layer should take the form
\begin{align}
\psi_0(\rb,t)=\bigg[\ee^{-\ii q_{0z}z} + \sum_{\Gb,\pm} B^\pm_{\Gb}\; \ee^{\ii(\Gb\cdot\Rb\pm q_{\Gb\varepsilon_0 z}z)} \;\Theta{(\pm z)}\bigg]\;\ee^{\ii(\q0parb\cdot\Rb-\varepsilon_0t)}, \label{psiLEED}
\end{align}
where $B^\pm_{\Gb}$ are Bragg scattering coefficients.\\

To determine $B^\pm_{\Gb}$ in Eq.~(\ref{psiLEED}), we start by expressing the incident wave around each atomic position
$\Rb_\alpha$~\cite{P1974} as
\begin{subequations}
\begin{align}
\ee^{\ii\qb_0\cdot\rb}=\ee^{\ii\q0parb\cdot\Rb_\alpha}\sum_{L}\phi^\mathrm{inc}_L j_L\big[q_0(\rb-\Rb_\alpha)\big], \label{incwave}
\end{align}
where
\begin{align}
\phi^\mathrm{inc}_L=4\pi\,Y_L^*\big(\Omega_{\qb_0}\big) \label{phiinc}
\end{align}
\end{subequations}
are expansion coefficients involving spherical harmonics $Y_L$ labeled by the quantum numbers $L=(l,m)$, while $j_L(q_0 \rb)=\ii^l j_l(q_0r) Y_L\big(\Omega_{\rr}\big)$ are regular spherical waves with a radial dependence introduced through spherical Bessel functions $j_l$. Note that we separate the phase associated with the propagation of the incident wave to each atomic site as a global factor $\ee^{\ii\q0parb\cdot\Rb_\alpha}$ in Eq.~(\ref{incwave}). Incidentally, under normal incidence ($\qb_0=-\zz$) we have  $\phi^\mathrm{inc}_L=\delta_{m0}\,(-1)^l\sqrt{4\pi\,(2l+1)}$. In the far field ($q_0r\gg1$), regular waves have the asymptotic behavior
\[
j_l(q_0r)=\frac{1}{2}\Big[h^{(1)}_l(q_0r)+h^{(2)}_l(q_0r)\Big]\approx\frac{1}{2q_0r}\Big[\ii^{-(l+1)}\ee^{\ii q_0r}+\ii^{l+1}\ee^{-\ii q_0r}\Big],
\]
which corresponds to a combination of outgoing and incoming waves expressed in terms of the Hankel functions $h^{(1)}_l$ and $h^{(2)}_l$, respectively. Scattering by an atom only affects the outgoing wave, which retains its magnitude and quantum numbers $L$ because $V_{\rm atom}(r)$ is a real and isotropic potential. Therefore, the waves become
\[
j_l(q_0 r)\rightarrow\frac{1}{2q_0r}\Big[e^{2i\delta_l}h^{(1)}_l(q_0 r)+h^{(2)}_l(q_0 r)\Big],
\]
where $\delta_l$ are phase shifts that depend on the orbital quantum number $l$ and the atomic potential (i.e., the type of atom). Here, we calculate $\delta_l$ numerically assuming free atoms, for which the potential is obtained by using a Dirac-Fock code~\cite{AZR96}, while electron scattering is computed by solving the Schr\"odinger equation following a well-established procedure~\cite{SM91_2}. For an individual atom, the scattered components can thus be written as~\cite{paper031} $\sum_{L}\phi^\mathrm{scat}_L h^{(+)}_L\big[(q_0 (\rb-\Rb_\alpha)\big]$, where $h^{(+)}_L(q_0 \rb)=\ii^{l+1} h^{(1)}_l(q_0 r) Y_L\big(\Omega_{\rr}\big)$ are spherical outgoing waves and $\phi^\mathrm{scat}_L=t_l\,\phi^\mathrm{inc}_L$ are scattering coefficients that involve the scattering matrix elements $t_l=\ee^{\ii\delta_l}\sin\delta_l$.

In the atomic monolayer, lattice symmetry allows us to write
\begin{align}
\psi_0(\rb,t)=\bigg[\ee^{\ii\qb_0\cdot\rb} + \sum_\alpha \ee^{\ii\q0parb\cdot\Rb_\alpha} \sum_L\phi^\mathrm{scat}_L\,h^{(+)}_L\big[(q_0 (\rb-\Rb_\alpha)\bigg]\;\ee^{-\ii \varepsilon_0 t}, \label{psiLEEDbis}
\end{align}
where the propagation phase factor $\ee^{\ii\q0parb\cdot\Rb_\alpha}$ is retained in the contribution of scattering by each lattice site $\alpha$. Now, we incorporate multiple scattering in the monolayer by supplementing the incident wave at each atom with the result of scattering from the rest of the atoms. More precisely, we expand the waves outgoing from each position $\beta$ as a combination of regular waves at each other position $\alpha$ using the identity~\cite{P1974} $h^{(+)}_{L'}\big[(q_0 (\rb-\Rb_\beta)\big]=\sum_L G_{\alpha\beta,LL'}\,j_L\big[(q_0 (\rb-\Rb_\alpha)\big]$, where $G_{\alpha\beta,LL'}=4\pi\sum_{L''}h^{(+)}_{L''}\big[(q_0 (\Rb_\alpha-\Rb_\beta)\big]\int\dd\Omega\;Y_L(\Omega)Y_{L''}(\Omega)Y_{L'}^*(\Omega)$. From these considerations, we find the self-consistent equation $\phi^\text{scat}_L=t_l\,\big[\phi^\text{inc}_L+\sum_{L'}\,G_{LL'}(\q0parb)\,\phi^\text{scat}_{L'}\big]$, or equivalently,
\begin{align}
\phi^\text{scat}_L=\sum_{L'}\big[S^{-1}\big]_{LL'} t_{l'}\,\phi^\text{inc}_{L'}, \label{phiscat}
\end{align}
where we define the lattice sum $G_{LL'}(\q0parb)=\sum_{\beta\neq\alpha}\,\ee^{\ii\q0parb\cdot(\Rb_\beta-\Rb_\alpha)}\,G_{\alpha\beta,LL'}$, which is trivially independent of $\alpha$, and the matrix $S_{LL'}=t_l\,G_{LL'}(\q0parb)$. In our simulations, we use the efficient numerical methods developed by Kambe~\cite{K1967,P1974} to carry out the sum over $\beta$ and find converged results by truncating the multipolar expansion to $l\le5$.

To bring Eq.~(\ref{psiLEEDbis}) into the form of Eq.~(\ref{psiLEED}), we use the identity (Eq.~(A.39) in Ref.~\cite{P1974})
\[
h_L^{(+)}(q_0\rb)=-\frac{1}{2\pi^2q_0}\int\dd^3\qb\;Y_L\big(\Omega_\qb\big)\,\frac{\ee^{\ii\qb\cdot\rb}}{q_0^2-q^2+\ii0^+}
\]
and then convert the sum over $\alpha$ into a sum over reciprocal lattice vectors as
\[
\sum_\alpha\ee^{\ii(\q0parb-\qparb)\cdot\Rb_\alpha}=\frac{(2\pi)^2}{A}\sum_\Gb\delta(\q0parb+\Gb-\qparb).
\]
The $\qparb$ integral is done using these $\delta$-functions, while the integral over $q_z$ can be performed by closing the integration contour in the upper (lower) complex $q_z$ plane for $z>0$ ($z<0$). Finally, we recover an expression like Eq.~(\ref{psiLEED}), from which we identify
\begin{align}
B^\pm_\Gb&=\frac{2\pi\ii}{A\,q_0q_{\Gb\varepsilon_0 z}} \sum_L Y_L\big(\Omega_{\q0parb+\Gb\pm q_{\Gb\varepsilon_0 z}\zz}\big)\,\phi_L^\mathrm{scat} =\frac{8\pi^2\ii}{A\,q_0q_{\Gb\varepsilon_0 z}} \sum_{LL'} Y_L\big(\Omega_{\q0parb+\Gb\pm q_{\Gb\varepsilon_0 z}\zz}\big)\,
\big[S^{-1}\big]_{LL'}\,t_{l'}\,Y_{L'}^*\big(\Omega_{\qb_0}\big), \label{BG}
\end{align}
where the rightmost identity is obtained by inserting the scattering wave coefficients $\phi_L^\mathrm{scat}$ obtained from Eqs.~(\ref{phiinc}) and (\ref{phiscat}).

From the above considerations, we can also obtain the Green function $\mathcal{G}_0(\rb,\rb',\varepsilon)$ corresponding to an electron energy $\hbar\varepsilon$. In analogy to Eq.~(\ref{eq:G3D_to_G1D}), we separate it into the contribution of different in-plane wave vector components $\qparb$, each of them connected to the whole set of in-plane wave vectors $\big\{\qparb+\Gb\big\}$ (differing by reciprocal lattice vectors $\Gb$) {\it via} scattering by the atomic monolayer. We thus restrict $\qparb$ to the first Brillouin zone (1BZ) and write the frequency-space Green function as
\begin{subequations}
\label{GrGG}
\begin{align}
\mathcal{G}_0(\rb,\rb',t-t')=\sum_{\Gb\Gb'}\int_{\rm 1BZ}\frac{\dd^2\qparb}{(2\pi)^2}\; \ee^{\ii(\qparb+\Gb)\cdot\Rb-\ii(\qparb+\Gb')\cdot\Rb'} \int\frac{\dd\varepsilon}{2\pi}\;\ee^{\ii\varepsilon(t'-t)}
\;\mathcal{G}_{\Gb\Gb'}\big(\qparb,z,z',\varepsilon\big).
\end{align}
Energy conservation in the regions outside the monolayer (i.e., above and below) implies that $\mathcal{G}_{\Gb\Gb'}\big(\qparb,z,z',\varepsilon\big)$ must be made of components with a $(z,z')$ dependence given by $\ee^{\pm\ii q_{\Gb\varepsilon z}z\pm\ii q_{\Gb'\varepsilon z}z'}$, where $q_{\Gb\varepsilon z}=\sqrt{2\me\varepsilon/\hbar-|\qparb+\Gb|^2+\ii0^+}$. Similarly to Eq.~(\ref{Gdelta}), we expect the Green function to consist of a free-space term plus contributions arising from wave scattering at the atomic layer (i.e., reflection and transmission), that is,
\begin{align}
\mathcal{G}_{\Gb\Gb'}\big(\qparb,z,z',\varepsilon\big)=-\frac{\ii\me}{\hbar^2q_{\Gb\varepsilon z}}\,\Big[ \delta_{\Gb\Gb'} \ee^{\ii q_{\Gb\varepsilon z}|z-z'|}+r^{ss'}_{\Gb\Gb'}(\qparb,\varepsilon)\,\ee^{\ii q_{\Gb\varepsilon z}|z|+\ii q_{\Gb'\varepsilon z}|z'|} \Big], \label{GGG}
\end{align}
where $r^{ss'}_{\Gb\Gb'}(\qparb,\varepsilon)$ are Bragg scattering coefficients with $s=\sign\{z\}$ and $s'=\sign\{z'\}$ denoting the direction of wave propagation (i.e., upward and downward for $s,s'=1$ and -1, respectively). Following the same steps as in the derivation presented above for Eqs.~(\ref{psiLEED}) and (\ref{BG}), but now applied to incident waves $\ee^{\ii(\qparb+\Gb')\cdot\Rb+\ii s'q_{\Gb'\varepsilon z}z}$ and outgoing waves $\ee^{\ii(\qparb+\Gb)\cdot\Rb+\ii sq_{\Gb\varepsilon,z}z}$ (with $s=\pm1$ and $s'=\pm1$ indicating upward/downward propagation), we can readily write
\begin{align}
r^{ss'}_{\Gb\Gb'}(\qparb,\varepsilon)=\frac{8\pi^2\ii}{A\,q_0q_{\Gb\varepsilon z}} \sum_{LL'} Y_L\big(\Omega_{\qparb+\Gb+sq_{\Gb\varepsilon z}\zz}\big)\,
\big[S^{-1}\big]_{LL'}\,t_{l'}\,Y_{L'}^*\big(\Omega_{\qparb+\Gb'+s'q_{\Gb'\varepsilon z}\zz}\big). \label{rGG}
\end{align}
\end{subequations}
We note that Eqs.~(\ref{GrGG}) are only valid outside the range of the atomic potentials $V_{\rm atom}(|\rb-\Rb_\alpha|)$, which represents a thin region and can therefore be safely neglected to describe the interaction with light via the LS equation [Eq.~(\ref{eq:LippSchwEq})].

\subsection{Interaction with light}

The Green function in Eqs.~(\ref{GrGG}) accounts for electron--lattice interaction to all orders of scattering. Inserting it into Eq.~(\ref{eq:LippSchwEq}) together with the initial wave function in Eq.~(\ref{psiLEED}), we can obtain a perturbative solution to the wave function up to an arbitrary order $n$ in the interaction with light. After a net exchange of $\ell$ photons (with $|\ell|\le n$), the in-plane wave vector and frequency of the electron become $\q0parb+\ell\kparb$ and $\varepsilon_\ell=\varepsilon_0+\ell\omega$, respectively, as deduced from the LS equation by performing the $\Rb'$ and $t'$ integrals, which then select the $\mathcal{G}_{\Gb\Gb'}\big(\q0parb+\ell\kparb,z,z',\varepsilon_0+\ell\omega\big)$ Green function component.

A general analysis is involved and does not provide substantial insights. Instead, we present results assuming normally incident electrons ($\q0parb=0$ and $\qb_0=-q_0\zz$) and limit our discussion to first-order interaction with a grazingly incident light plane wave of wave vector $\kparb=k\,\xx$ and real electric field amplitude $\Eb_0$ in the $y-z$ plane (i.e., $\Ab(\rb,t)=(\Eb_0/\ii k)[\ee^{\ii(kx-\omega t)}-\ee^{-\ii(kx-\omega t)}]$). Then, the $\ell=0$ wave function component remains as in Eq.~(\ref{psiLEED}), while two sidebands $\ell=\pm1$ emerge at order $n=1$. Inserting the  field into Eq.~(\ref{H1}) and combining the result with Eqs.~(\ref{eq:LippSchwEq}), (\ref{psiLEED}), and (\ref{GrGG}), we obtain
\begin{align}
\psi^{(1)}(\rb,t)=\frac{e}{\hbar\omega}&\sum_{\ell=\pm1} \ii^{\ell-1}\;\ee^{-\ii\varepsilon_\ell t}
\sum_{\Gb\Gb'} \frac{1}{q_{\Gb\varepsilon_\ell z}} \ee^{\ii(\ell\kparb+\Gb)\cdot\Rb}
\nonumber\\ &\times
\Eb_0\cdot
\sum_{\pm}\int\dd z'\;\Theta{(\pm z')}
\;\ee^{-\gamma |z'|}\;
\Big[ \delta_{\Gb\Gb'} \ee^{\ii q_{\Gb\varepsilon_\ell z}|z-z'|}+r^{\sign\{z\},\pm}_{\Gb\Gb'}(\ell\kparb,\varepsilon_\ell)\,\ee^{\ii q_{\Gb\varepsilon_\ell z}|z|+\ii q_{\Gb'\varepsilon_\ell z}|z'|} \Big]
\nonumber\\ & \quad\quad\quad\quad\quad\quad\quad\quad\quad\quad\quad\;\,\times
\Big[
-\zz\,q_0\,\delta_{\Gb'0}\;\ee^{-\ii q_0z'}
+ B^\pm_{\Gb'}\;(\Gb'\pm q_{\Gb'\varepsilon_0z}\zz)
\;\ee^{\ii q_{\Gb'\varepsilon_0z}|z'|} \Big], \nonumber
\end{align}
where we have inserted the $\ee^{-\gamma |z'|}$ factor that we discuss in Sec.~\ref{rrecursive}. Finally, performing the $z'$ integral with the help of Eqs.~(\ref{integrals}) and taking the $|z|\to\infty$ and $\gamma\to0^+$ limits in that order, we obtain
\begin{align}
&\psi^{(1)}(\rb,t)=\sum_{\Gb,\ell,\pm}
C_{\Gb,\ell}^\pm \;
\ee^{\ii(\ell\kparb+\Gb)\cdot\Rb\pm\ii q_{\Gb\varepsilon_\ell z}z-\ii\varepsilon_\ell t}
\;\Theta(\pm z), \nonumber
\end{align}
where
\begin{align}
C_{\Gb,\ell}^s=&-\frac{\ii^\ell e}{\hbar\omega} \;
\frac{1}{q_{\Gb\varepsilon_\ell z}}  \Eb_0\cdot\sum_{\pm} 
\Bigg\{
\zz\,r^{s\pm}_{\Gb0}(\ell\kparb,\varepsilon_\ell)\,\frac{q_0}{q_{\Gb\varepsilon_\ell z}\mp q_0}
\label{S23} \\
&+\sum_{\Gb'} B^\pm_{\Gb'}\;(\Gb'\pm q_{\Gb'\varepsilon_0z}\zz)
\bigg[\delta_{\Gb\Gb'}
\frac{\pm s}{q_{\Gb\varepsilon_\ell z}\mp sq_{\Gb'\varepsilon_0z}}
-r^{s\pm}_{\Gb\Gb'}(\ell\kparb,\varepsilon_\ell)\,\frac{1}{q_{\Gb\varepsilon_\ell z}+q_{\Gb'\varepsilon_0z}} \bigg]
\Bigg\} \nonumber
\end{align}
are the sought-after inelastic scattering coefficients.

\end{widetext}


%

\pagebreak
\renewcommand{\thefigure}{S\arabic{figure}} 

\begin{figure*}
\centering\includegraphics[width=0.74\textwidth]{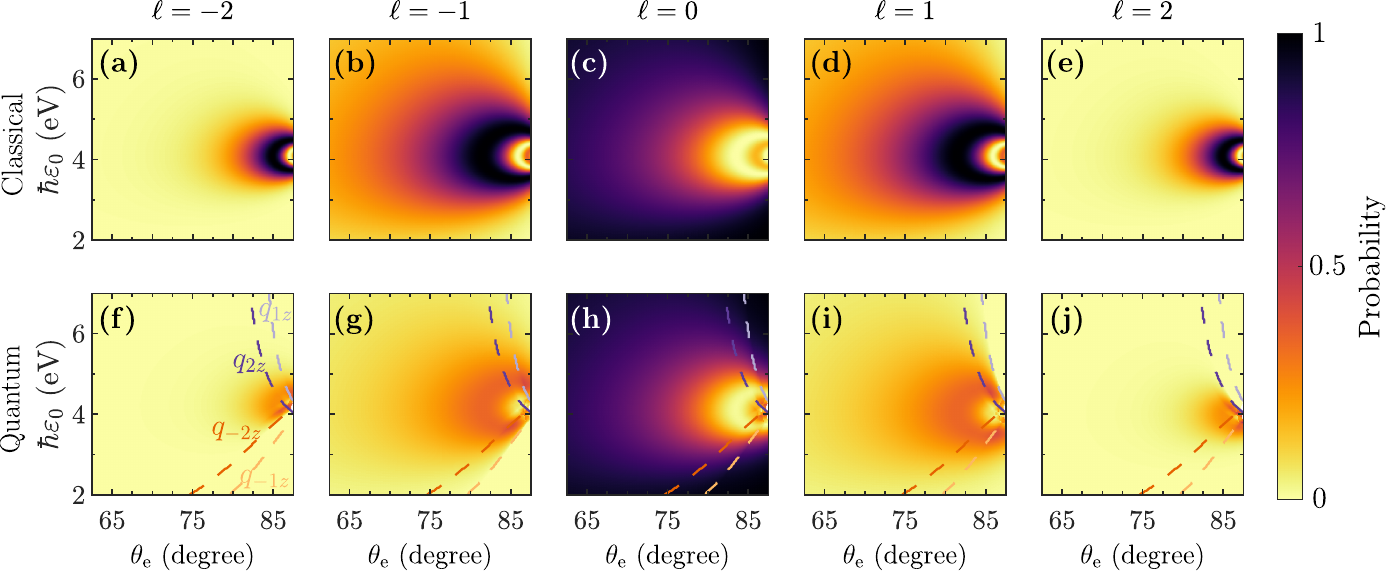}
\caption{\textbf{Inelastic scattering of low-energy electrons upon total reflection at a polariton-supporting surface.} Panels (b-d,g-i) are reproduced from Fig.~2a-c,e-g in the main text, but now including dashed curves indicating the kinematic threshold beyond which energy--momentum conservation renders $q_{\ell z}$ purely imaginary (i.e., $1+\ell \omega/\varepsilon_0=|\sin\thetae+\ell \kpar/q_0|^2$). In addition, we present results for $\ell=\pm2$ in panels (a,e,f,j).}
\label{FigS1}
\end{figure*}

\begin{figure*}
\centering\includegraphics[width=0.74\textwidth]{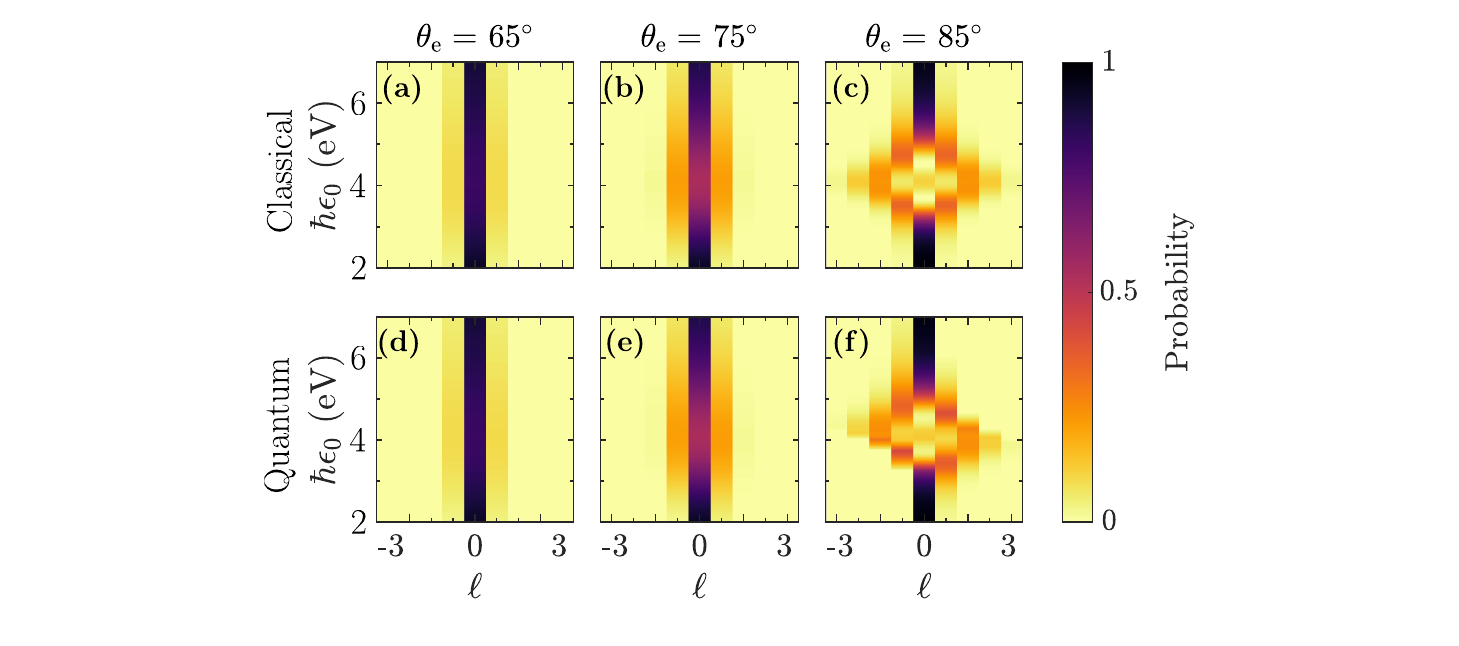}
\caption{\textbf{Influence of electron incidence angle on inelastic scattering under the conditions of Fig.~2 in the main text.} We present results analogous to Fig.~2d,h, but for different values of the electron incidence angle $\thetae$.
}
\label{FigS2}
\end{figure*}

\begin{figure*}
\centering\includegraphics[width=0.85\textwidth]{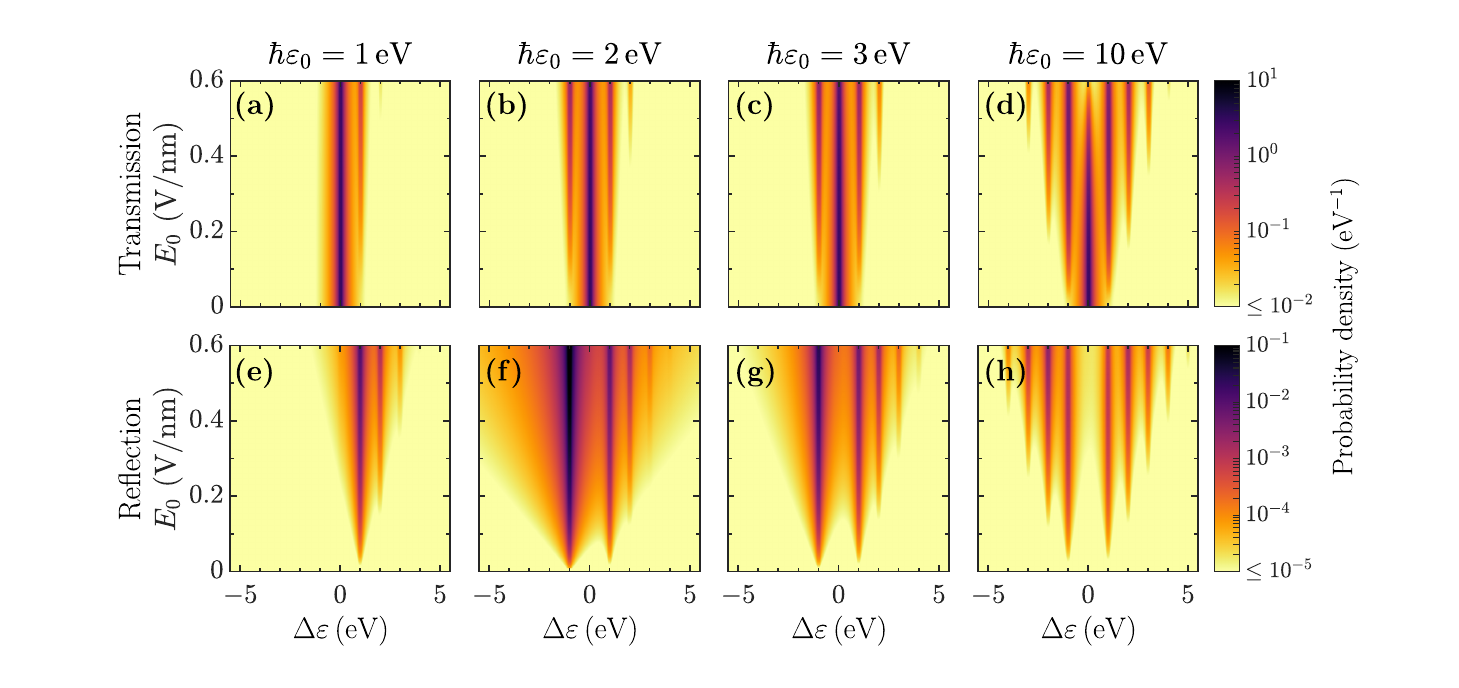}
\caption{\textbf{Inelastic scattering of low-energy electrons at an electron-transparent, polariton-supporting thin film.} Same as Fig.~3 in the main text, but with $U_0=0$ (i.e., the electron does not see the material).
}
\label{FigS3}
\end{figure*}

\end{document}